\documentstyle[preprint,aps,floats,epsf,tighten]{revtex}

\def\aprle{\buildrel < \over {_{\sim}}}

\begin{document}
\begin{flushright}
	FERMILAB-Pub-01/228-T\\
	hep-ph/0108070\\
	October 2001
\end{flushright}
\vspace*{0.75in}
\begin{center}
{\large\bf GUT Model Predictions for Neutrino Oscillation 
Parameters Compatible with the Large Mixing Angle Solar Solution}\\[1in]
Carl H. Albright$^{1,2}$ and S. Geer$^2$\\[0.5in]
\it
$^1$ Department of Physics, Northern Illinois University, DeKalb, IL 60115\\
$^2$Fermi National Accelerator Laboratory, P.O. Box 500,
Batavia, IL 60510, USA\\[0.75in]

\end{center}
\thispagestyle{empty}

\begin{abstract}

Within the framework of an SO(10) GUT model that can accommodate both the 
atmospheric and the LMA solar neutrino mixing solutions, we present explicit 
predictions for the neutrino 
oscillation parameters $\sin^2 2\theta_{13}$, $\sin^2 2\theta_{12}$, 
$\sin^2 2\theta_{23}$, and $\Delta m^2_{21}$.  Precise measurements of 
$\sin^2 2\theta_{12}$ and $\Delta m^2_{21}$ by KamLAND can be used to 
precisely determine the GUT model parameters.  We find that the model can 
then be tested at Neutrino Superbeams and Neutrino Factories with precision 
neutrino oscillation measurements of $\sin^2 2\theta_{23},\ 
\sin^2 2\theta_{13}$, and the leptonic CP phase $\delta_{CP}$.
\end{abstract}
\vspace*{0.75in}

\vspace*{0.25in}

Electronic addresses: albright@fnal.gov, sgeer@fnal.gov

\newpage

\section{Introduction}

Over the last few years the evidence for neutrino oscillations between
the three known active-neutrino flavors
($\nu_e, \nu_\mu$, and $\nu_\tau$) has become increasingly convincing. The
atmospheric neutrino flux measurements from
the Super-Kamiokande (Super-K) experiment exhibit a deficit of muon neutrinos
which varies with zenith angle (and hence baseline) in a way consistent with
$\nu_\mu \to \nu_x$ oscillations \cite{atm}. In principle $\nu_x$ could be 
$\nu_e, \nu_\tau$, $\nu_s$ (where $\nu_s$ is a light sterile neutrino), or some
combination of these. However, further Super-K measurements exclude
$\nu_x$ being predominantly $\nu_s$, and reactor $\nu_e$ disappearance results
from the CHOOZ experiment \cite{CHOOZ} exclude $\nu_x$ being predominantly 
$\nu_e$. Hence, the Super-K atmospheric neutrino measurements provide
strong evidence for $\nu_\mu \to \nu_\tau$ oscillations; indeed there is  some 
evidence for $\nu_\tau$ interactions in the Super-K data. In addition to
the atmospheric neutrino deficit, there has been the long-standing result,
first obtained from the Homestake experiment \cite{homestake}, that the 
$\nu_e$ flux from
the sun is less than expected. The recent measurement of the total flux of
active neutrinos from the sun obtained from the SNO experiment \cite{SNO} is
consistent with the predicted flux from solar models \cite{Bahcall}. Hence, 
when taken together with solar neutrino measurements from Super-K 
\cite{solSK}, the SNO results
imply that there is a component of active neutrinos within the solar flux
that is not $\nu_e$, and hence that $\nu_e \to \nu_x$ oscillations are
taking place, where $\nu_x$ can be $\nu_\mu$ and/or $\nu_\tau$. The solar
neutrino and
atmospheric neutrino results, taken together, suggest that oscillations
occur between all three known active flavors.

The atmospheric neutrino data are consistent with $\nu_\mu \to \nu_\tau$
oscillations provided the oscillation parameters that define the
oscillation amplitude and frequency lie in one well-defined region
of parameter space. In contrast, the solar neutrino measurements are
currently consistent with the associated oscillation parameters
being within any of four regions of parameter space. However,
although the evidence is not yet compelling, the data
seem to exhibit a preference for one of these regions of parameter space,
namely the one corresponding to the Large Mixing Angle (LMA) MSW solution
\cite{MSW}.

The splittings between the squares of the masses of the neutrino mass
eigenstates determine the oscillation frequency.
The atmospheric- and solar-neutrino oscillation data imply that neutrinos
have masses in the range $10^{-5} - 1$ eV. This mass scale can be accommodated 
naturally within the framework of models based on Grand Unified Theories 
(GUTs). The very small neutrino mass is easily generated by the seesaw 
mechanism \cite{gmrsy} in which the light neutrino mass matrix is obtained 
from the Dirac and right-handed Majorana neutrino mass matrices.

Grand Unified models provide a theory of flavor, and relate quark masses and 
mixings to lepton masses and mixings. Hence, neutrino oscillation data,
which measure neutrino masses and mixings, constrain GUT models.
In this paper, for one promising GUT model, we explore how future
neutrino oscillation experiments can test the theory. We restrict ourselves
to the LMA solution for the solar neutrino data, and provide predictions
for the neutrino mass-splittings and mixing angles that will be measured
in the next few years.

\section{Three-Flavor Mixing}
\label{sec:theory}

Within the framework of three-flavor mixing, 
the flavor eigenstates $\nu_\alpha\ (\alpha = e, \mu, \tau)$ are related to
the mass eigenstates $\nu_j\ (j = 1, 2, 3)$ in vacuum by
\begin{equation}
\nu_\alpha = \sum_j U_{\alpha j} \nu_j \;,
\end{equation}
where $U$ is the unitary $3\times3$ Maki-Nakagawa-Sakata (MNS) mixing matrix
\cite{MNS} times a diagonal phase matrix $\Phi_M$: $U = U_{MNS}\Phi_M$. 
The MNS mixing matrix is conventionally specified by 3 mixing angles 
($\theta_{23}, \theta_{12}, \theta_{13}$) and a CP-violating phase
($\delta_{CP}$) with the parameterization
\begin{equation}
  U_{MNS} = \left(\matrix{c_{12}c_{13} & s_{12}c_{13} & s_{13}e^{-i\delta_{CP}}
	\cr  -s_{12}c_{23} - c_{12}s_{23}s_{13}e^{i\delta_{CP}} &
        c_{12}c_{23} - s_{12}s_{23}s_{13}e^{i\delta_{CP}} & s_{23}c_{13}\cr
        s_{12}s_{23} - c_{12}c_{23}s_{13}e^{i\delta_{CP}} &
        -c_{12}s_{23} - s_{12}c_{23}s_{13}e^{i\delta_{CP}} & c_{23}c_{13}
        \cr}\right),
\label{eq:mns}
\end{equation}
where $c_{jk} \equiv \cos\theta_{jk}$ and $s_{jk} \equiv
\sin\theta_{jk}$.  The angles can be restricted to the first quadrant,
$0\le \theta_{ij} \le \pi/2$,
with $\delta_{CP}$ in the range $-\pi \le \delta_{CP} \le \pi$, though it will later
prove advantageous to consider $\theta_{13}$ in the fourth quadrant.  The 
$\Phi_M$ phase matrix has the form 
\begin{equation}
  \Phi_M = {\rm diag} (e^{i\chi_1},\ e^{i\chi_2},\ 1),
\label{eq:majphase}
\end{equation}
where $\chi_1$ and $\chi_2$ are Majorana phases which can not be rotated 
away.

The atmospheric neutrino oscillation data indicate that \cite{atm}
\begin{equation}
\begin{array}{rcl}
        |\Delta m^2_{32}| & \simeq& 3.2 \times 10^{-3}\ {\rm eV^2},\\[4pt]
        \sin^2 2\theta_{23} & =& 1.0,\ (\geq 0.89 \ {\rm at}\ 90\%\ 
		{\rm c.l.}),\\
\end{array}\label{eq:atm}
\end{equation}
where $\Delta m^2_{ij} \equiv m^2_i - m^2_j$ and $m_1,\ m_2$ and $m_3$ are
the mass eigenstates..
The atmospheric neutrino oscillation amplitude can be expressed in terms of 
the $U_{MNS}$ matrix elements and is given by
$\sin^2 2\theta_{atm} = 4|U_{\mu 3}|^2(1 - |U_{\mu 3}|^2) \simeq 
4|U_{\mu 3}|^2 |U_{\tau 3}|^2$.  The approximation is valid because 
$|U_{e3}|$ is known to be small \cite{CHOOZ}.

The solar neutrino oscillation data from Super-K indicate that,
for the LMA solution, the allowed region is approximately bounded by 
\begin{equation}
\begin{array}{rcl}
        \Delta m^2_{21} & =& (2.2 - 17) \times 10^{-5}\ {\rm eV^2},\\[4pt]
        \sin^2 2\theta_{sol} & =& (0.6 - 0.9),\\
\end{array}\label{eq:sol}
\end{equation}
where the solar neutrino oscillation amplitude is given by
$\sin^2 2\theta_{sol} = 4|U_{e1}|^2 (1 - |U_{e1}|^2) \simeq 
4|U_{e1}|^2 |U_{e2}|^2$.  In defining
the viable region of GUT model parameter space we shall make use of the
allowed LMA solar mixing region specified in~\cite{solSK}.  Other
recent analyses also prefer the LMA solution \cite{LMA} .
\newpage
\section{The GUT Model}

The GUT model which shall be studied here was developed by Albright and Barr 
\cite{ab} and is based on the grand unified group $SO(10)$ with a $U(1) \times 
Z_2 \times Z_2$ flavor symmetry.  We adopt this model in our present study 
because it can accommodate the LMA solution and makes quantitative predictions
for the measured oscillation parameters.
The model involves a minimum set of Higgs fields 
which solves the doublet-triplet
splitting problem.  This requires just one ${\bf 45}_H$ whose VEV points in 
the $B-L$ direction, and there are no higher rank representations.  Two pairs 
of ${\bf 16}_H,\ {\bf \overline{16}}_H$'s stabilize the solution \cite{br}.
Several Higgs in the ${\bf 10}_H$ representations, together with Higgs singlets,
are also present.  The Higgs superpotential exhibits the $U(1) \times Z_2 
\times Z_2$ symmetry \cite{br} which is used for the flavor symmetry 
of the GUT model.  The combination of VEVs, $\langle {\bf 45_H}\rangle_{B-L},
\ \langle 1({\bf 16_H})\rangle$ and $\ \langle 1({\bf \overline{16}_H})\rangle$
break $SO(10)$ to the Standard Model.  The electroweak VEVs arise from the 
combinations $v_u = \langle 5({\bf 10_H})\rangle$ and $v_d = \langle 
\overline{5}({\bf 10_H})
\rangle\cos \gamma + \langle \overline{5}({\bf 16'_H})\rangle \sin \gamma$,
while the combination orthogonal to $v_d$ gets massive at the GUT scale.
As such, Yukawa coupling unification can be achieved at the GUT scale with
$\tan \beta \sim 2 - 55$, depending upon the $\overline{5}({\bf 10_H}) -
\overline{5}({\bf 16_H})$ mixing present for the $v_d$ VEV.
In addition, matter superfields appear in the following representations:
${\bf 16_1},\ {\bf 16_2},\ {\bf 16_3};\ {\bf 16},\ {\bf \overline{16}},
\ {\bf 16'}$, ${\bf \overline{16'}},\ {\bf 10_1},\ {\bf 10_2}$, and ${\bf 1}$'s,
where all but the ${\bf 16_i}\ (i = 1,2,3)$ get superheavy and are integrated
out.

The Dirac mass matrices for the up quarks, down quarks, neutrinos and charged 
leptons are found to be\\[-0.2in]
\begin{equation}
\begin{array}{ll}
U = \left(\matrix{ \eta & 0 & 0 \cr
  0 & 0 & \epsilon/3 \cr 0 & - \epsilon/3 & 1\cr} \right)M_U,\
  & D = \left(\matrix{ 0 & \delta & \delta' e^{i\phi}\cr
  \delta & 0 & \sigma + \epsilon/3  \cr
  \delta' e^{i \phi} & - \epsilon/3 & 1\cr} \right)M_D, \\[0.5in]
N = \left(\matrix{ \eta & 0 & 0 \cr 0 & 0 & - \epsilon \cr
        0 & \epsilon & 1\cr} \right)M_U,\
  & L = \left(\matrix{ 0 & \delta & \delta' e^{i \phi} \cr
  \delta & 0 & -\epsilon \cr \delta' e^{i\phi} &
  \sigma + \epsilon & 1\cr} \right)M_D,\\
\end{array}\label{eq:Dirac}
\end{equation}
\vspace*{-0.1in}
where
\vspace*{-0.1in}
\begin{equation}
\begin{array}{rlrl}
        M_U&\simeq 113\ {\rm GeV},&\qquad M_D&\simeq 1\ {\rm GeV},\\
        \sigma&=1.78,&\qquad \epsilon&=0.145,\\
        \delta&=0.0086,&\qquad \delta'&= 0.0079,\\
        \phi&= 126^\circ,& \qquad \eta&= 8 \times 10^{-6}\\
\end{array}\label{eq:input}
\end{equation}
are input parameters defined at the GUT scale to fit the low scale
observables after evolution downward from $\Lambda_{GUT}$.
Note that the phase $\phi$ was incorrectly stated as $54^\circ$ in \cite{ab}.
The above textures were obtained by imposing the Georgi-Jarlskog relations
\cite{gj} at $\Lambda_{GUT}$, $m^0_s \simeq m^0_\mu/3,\ m^0_d \simeq 3m^0_e$
with Yukawa coupling unification holding for $\tan \beta \sim 5$.
The matrix element contributions can be understood in terms of
Froggatt-Nielsen diagrams \cite{fn} as explained in \cite{ab}.  

All nine quark and charged lepton masses, plus the three CKM angles and CP
phase, are well-fitted with the eight input parameters.  With no extra 
phases present, aside from the one appearing in the CKM mixing matrix, the 
vertex of the
CKM unitary triangle occurs at the center of the presently allowed region
with $\sin 2\beta \simeq 0.64$.  The Hermitian matrices $U^\dagger U,\
D^\dagger D$, and $N^\dagger N$ are diagonalized with small left-handed
rotations, while $L^\dagger L$ is diagonalized by a large left-handed rotation.
This accounts for the small value of $V_{cb} = (U^\dagger_U U_D)_{cb}$,
while $|U_{\mu 3}| = |(U^\dagger_L U_\nu)_{\mu 3}|$ will turn out to be 
large for any reasonable right-handed Majorana mass matrix, $M_R$ \cite{abb}.

The effective light neutrino mass matrix, $M_\nu$, is obtained from the seesaw 
mechanism \cite{gmrsy} whereby $M_\nu = N^T M_R^{-1} N$. 
While the large atmospheric neutrino mixing $\nu_\mu \leftrightarrow \nu_\tau$
arises primarily from the structure of the charged lepton mass matrix, 
the solar and atmospheric mixings are essentially decoupled in the
model, so the structure of the right-handed Majorana mass matrix determines the
type of $\nu_e \leftrightarrow \nu_\mu,\ \nu_\tau$ solar neutrino mixing.
Any one of the recently favored four solar neutrino mixing solutions
can be obtained.  The LMA solution relevant to our study here requires some
fine-tuning and a hierarchical structure, but this can be explained in 
terms of Froggatt-Nielsen diagrams.  The most general form 
for the right-handed Majorana mass matrix we consider is \cite{ab}
\begin{equation}
          M_R = \left(\matrix{c^2 \eta^2 & -b\epsilon\eta & a\eta\cr
                -b\epsilon\eta & \epsilon^2 & -\epsilon\cr
                a\eta & -\epsilon & 1\cr}\right)\Lambda_R,\\
\label{eq:Maj}
\end{equation}
where the parameters $\epsilon$ and $\eta$ are those introduced in 
Eq.(\ref{eq:Dirac}) for the Dirac sector.  
Note that the 2-3 subsector has zero determinant and is closely related to 
that of $N$, as can also be understood in terms of Froggatt-Nielsen diagrams. 
If we set $a = b = c$, there is just one hierachy present involving one
Higgs singlet which induces a $\Delta L = 2$ transition.  In this case 
the determinant of $M_R$ vanishes.  In order to have an invertible $M_R$ and a 
viable seesaw mechanism, for simplicity we set $b=c$ but choose $a \neq b$.
This is neatly explained in terms of two Higgs singlets which break lepton 
number.  One singlet contributes to all nine matrix elements while, by virtue 
of its flavor charge assignment, the other singlet modifies only the 13 and 
31 elements of $M_R$.

To obtain $U_{MNS}$ from the mass matrices $L$ and $M_\nu$, we compute
the unitary transformations $U_L$ and $U_\nu$ that diagonalize $L^\dagger L$
and $M^\dagger_\nu M_\nu$ and yield the squares of the charged and neutral
lepton mass eigenvalues, respectively.  Three arbitrary phase transformations 
can be performed on the columns of $U_L$ which are constructed from the 
eigenvectors of $L^\dagger L$.  However, since $M_\nu$ is complex symmetric, 
it can also be diagonalized by use of the same $U_\nu$: 
\begin{equation}
	U^T_\nu M_\nu U_\nu = {\rm diag} (m_1,\ -m_2,\ m_3).
\label{eq:majdiag}
\end{equation}
Since we want the light neutrino masses to be real, $U_\nu$ can not be 
arbitrarily phase transformed and is uniquely specified up to 
sign changes on its column eigenvectors.  The unitary mixing matrix $U$ in 
Eq. (1) is then given by 
\begin{equation}
	U = U_{MNS}\Phi_M = \left(\Phi^\dagger_{row} U^\dagger_L U_\nu 
		\Phi_{col}\right)\Phi_{col}^\dagger,
\label{eq:MNStrans}
\end{equation}
where $\Phi_{row}$ and $\Phi_{column}$ are the row and column phase 
transformations
\begin{equation}
\begin{array}{rl}
	\Phi_{row}&= diag (e^{-i\phi_1},\ e^{-i\phi_2},\ e^{-i\phi_3}),\\[0.1in]
	\Phi_{col}&= diag (e^{-i\chi_1},\ e^{-i\chi_2},\ 1)\\
\end{array}
\label{eq:phases}
\end{equation}
of $U^\dagger_L U_\nu$ needed to bring $U_{MNS}$ into the parametric form
of Eq.(\ref{eq:mns}) whereby the $e1,\ e2,\ \mu 3$ and $\tau 3$ elements 
are real and positive, the real parts of $\mu 2$ and $\tau 1$ are positive,
while the real parts of $\mu 1$ and $\tau 2$ are negative.  The last factor
$\Phi^\dagger_{col}$ serves to undo the column phase transformation on 
$U_\nu$ and is just the Majorana phase matrix, $\Phi_M = \Phi^\dagger_{col}$,
from which the two Majorana phases $\chi_1$ and $\chi_2$ can be extracted.
As noted above, one is 
free to replace individually the column vectors of $M_\nu$ by their 
negatives, so the Majorana phases have a $180^\circ$ ambiguity.  Finally, the 
leptonic CP phase $\delta_{CP}$ can be identified from the $e3$ element of 
$U_{MNS}$ or alternatively by constructing the Jarlskog invariant 
\cite{jarlskog}, $J = Im (U_{e2}U^*_{e3}U^*_{\mu 2}U_{\mu 3})$, of the 
untransformed $U^\dagger _L U_\nu$ matrix.  The quadrant in which the phase 
$\delta_{CP}$ lies is uniquely determined once the sign of $\sin \theta_{13}$ 
is specified.  In carrying out the phase transformations, we have reduced the 
six inherent phase factors in $U^\dagger_L U_\nu$ to just three physical ones,
$\phi_{CP},\ \chi_1$ and $\chi_2$.
 
As an example, with $a=1,\ b=c=2$ and $\Lambda_R = 2.4 \times 10^{14}$ GeV,
the seesaw mechanism results in the light neutrino mass matrix 
        \begin{equation}
          M_\nu = N^T M^{-1}_R N = \left(\matrix{ 0 & -\epsilon & 0\cr
                        -\epsilon & 0 & 2\epsilon\cr 0 & 2\epsilon & 1\cr}
                        \right)M^2_U/\Lambda_R\\
	\label{eq:lightex}
        \end{equation}
with three texture zeros.  We obtain
        \begin{equation}
        \begin{array}{ll}
	  \multicolumn{2}{l}{m_1 = 5.6 \times 10^{-3},\quad m_2 = 9.8 \times 
		10^{-3},\quad m_3 = 57 \times 10^{-3}\ {\rm eV},}\\
          M_1 = M_2 = 2.8 \times 10^{8}\ {\rm GeV},\quad & M_3 = 2.5
                \times 10^{14}\ {\rm GeV},\\
          \Delta m^2_{32} = 3.2 \times 10^{-3}\ {\rm eV^2},\quad &
                \sin^2 2\theta_{\rm atm} = 0.994,\\
          \Delta m^2_{21} = 6.5 \times 10^{-5}\ {\rm eV^2},
                \quad & \sin^2 2\theta_{\rm sol} = 0.88,\\
          U_{e3} = -0.01395-0.00085i,\quad &\sin^2 2\theta_{\rm reac} = 
		0.0008.\\
	  J = 2.0 \times 10^{-4},\quad \delta_{CP} = -3.5^\circ,\quad & 
		\chi_1 = -0.2^\circ, \quad \chi_2 = 0.1^\circ.\\	
        \end{array}\label{eq:lightexres}
        \end{equation}
Here we have chosen the convention $\sin \theta_{13} < 0$, so that the CP 
phase $\delta_{CP}$ is near zero rather than $180^\circ$. 
The small value of $\delta_{CP}$ follows since $M_\nu$ is real in this example,
while $L$ contributes only a small phase contribution.  The two Majorana
phases are very small, since essentially no phase rotation on the right
is needed to bring $U^\dagger_L U_\nu$ into the standard MNS form of 
Eq. (\ref{eq:mns}).  The effective neutrinoless double beta decay mass 
is given by 
\begin{equation}
\begin{array}{rl}
	\langle m_{\beta\beta} \rangle&= |\sum_i m_i U_{ei}^2| 
			= 5.7 \times 10^{-4}\ {\rm eV},\\
\end{array}
\label{eq:betabeta}
\end{equation}
where the Majorana phases and the signs of the eigenvalues in Eq. 
(\ref{eq:majdiag}) are taken into account.  In the GUT model we are considering,
$\langle m_{\beta\beta} \rangle \sim {\rm few} \times 10^{-4}$ eV is obtained 
over the entire viable LMA region.  Note that these values will not be 
accessible to the presently planned double beta decay experiments.  

The above results compare favorably with the determination of the 
atmospheric neutrino mixing parameters by the Super-K collaboration 
as given in Eq. (\ref{eq:atm}), as well 
as their present best-fit point in the solar neutrino LMA region as given in 
Eq. (\ref{eq:sol}).  In fact, the whole
presently-allowed LMA region \cite{LMA} can be covered with $1.0 \aprle a 
\aprle 2.4$ and $1.8 \aprle b=c \aprle 5.2$.

\section{Results}

We can now examine the viable region of GUT model parameter space that is
consistent with the LMA solar neutrino solution and explore the predicted
relationships between the observables
$\sin^2 2\theta_{23}, \sin^2 2\theta_{12}, \sin^2 2\theta_{13}, \delta_{CP}$,
$\Delta m^2_{32}$, and $\Delta m^2_{21}$. We will first consider the
simplest case in which there are, in effect, only two real dimensionless GUT
model parameters. We then look at the more general case in which we allow a
finite phase $\phi'$ so that $a$ is complex.
\vfill
\begin{figure}[h]
\centering\leavevmode
\epsfxsize=5.0in\epsffile{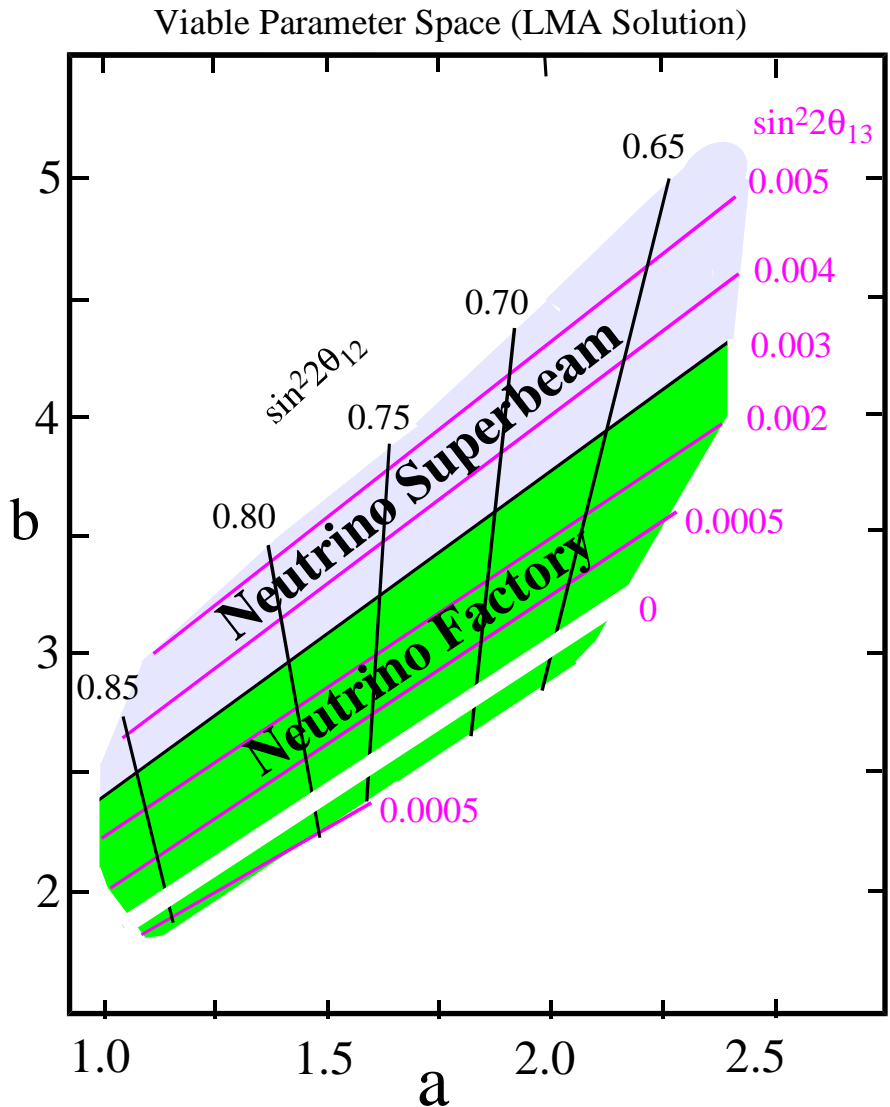}
%
%\medskip
%\vspace{0.2cm}
\caption[]{The viable region of GUT parameter space consistent 
with the present bounds on the LMA MSW solution. Contours of 
constant $\sin^2 2\theta_{13}$ and lines of constant 
$\sin^2 2\theta_{12}$ are shown. The region above 
$\sin^2 2\theta_{13} = 0.003$ can be explored with 
Neutrino Superbeams, while the region below this can be explored 
with Neutrino Factories, down to $\sin^2 2\theta_{13} \sim 0.0001$.}
\label{fig:gut_plot}
\end{figure}
\newpage
\subsection{Parameter Choice: $a$ and $b=c$ Real}

The viable region of GUT model parameter space consistent with the LMA solar
solution is shown in Fig.~\ref{fig:gut_plot}. Both parameters $a$ and $b$ are
constrained by the data to be close to unity, with $1.0 \aprle a \aprle 2.4$ 
and $1.8 \aprle b \aprle 5.2$. Superimposed on the allowed region,
Fig.~\ref{fig:gut_plot} shows contours of constant $\sin^2 2\theta_{12}$
(which are approximately parallel to the $b$-axis) and contours of constant
$\sin^2 2\theta_{13}$ (which are approximately at $45^\circ$ in the
$(a,b)$-plane).

\vspace*{0.1in}
\begin{figure}[h]
\centering\leavevmode
\epsfxsize=5.0in\epsffile{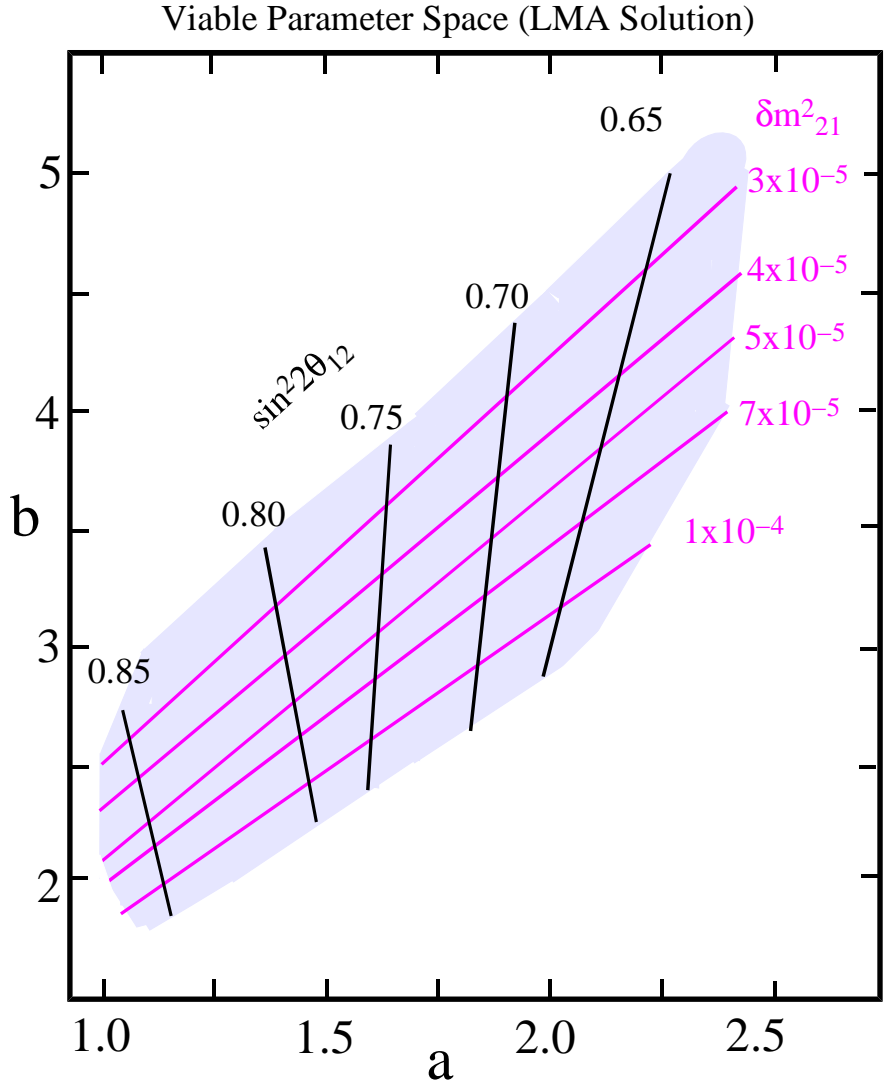}
\vspace{0.2cm}
\caption[]{The viable region of GUT parameter space consistent with the
present bounds on the LMA MSW solution. Contours of constant $\Delta m^2_{21}$
and lines of constant $\sin^2 2\theta_{12}$ are shown.}
\label{fig:gut_plot2}
\end{figure}
The coming long-baseline accelerator neutrino oscillation experiment MINOS 
\cite{MINOS} at Fermilab, and the CNGS experiments \cite{CNGS} at CERN, are 
expected
to be able to observe a $\nu_\mu \to \nu_e$ signal if $\sin^2 2\theta_{13} >
0.03$. This is above the allowed region of the $(a,b)$-parameter space. Hence 
the GUT model we are considering predicts that these long-baseline experiments
will obtain a null result.  A new generation of upgraded conventional neutrino 
beams 
is being considered \cite{superbeams},
and is expected to be able to probe the region $\sin^2 2\theta_{13} > 0.003$,
and hence measure the parameter $\theta_{13}$ if the solution lies in the
upper part of the allowed $(a,b)$-plane indicated in the figure. A Neutrino
Factory \cite{NF} is expected to be able to probe down to values of $\sin^2
2\theta_{13}$ as low as $O(10^{-4})$, which will therefore cover the entire
allowed $(a,b)$-plane, except for a narrow band in which $\sin^2 2\theta_{13}
\to 0$ as $\sin^2 2\theta_{23}$ becomes maximal. 

\begin{figure}[h]
\centering\leavevmode
\epsfxsize=5.0in\epsffile{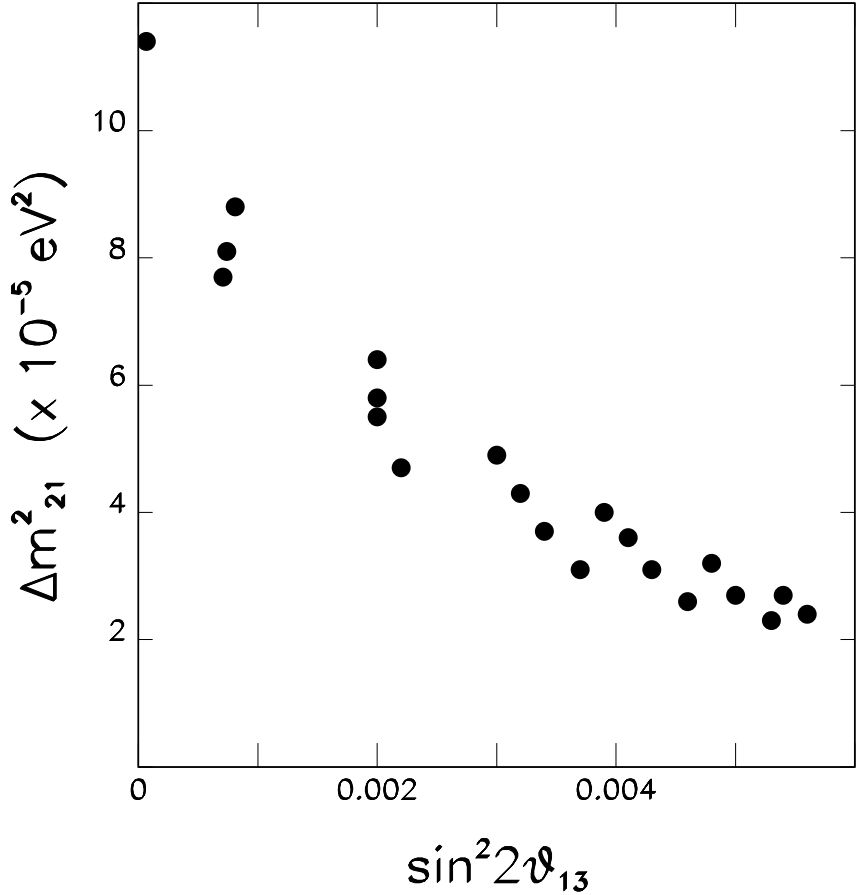}
\vspace{0.5cm}
\caption[]{Variation of $\sin^2 2\theta_{13}$ with $\Delta m^2_{21}$. The
points plotted populate a grid which spans the viable region of the
$(a,b)$ parameter space.  The small spread in points
across the band indicated arises from the variation in $\sin^2 2\theta_{12}$
for the points plotted.}
\label{fig:theta13_dm21}
\end{figure}

Figure~\ref{fig:gut_plot2} shows, once again, the viable region of parameter
space consistent with the LMA solar solution, but this time with contours of
constant $\Delta m^2_{21}$ displayed. These contours are approximately at
$45^\circ$ in the $(a,b)$-plane, and are almost parallel to the contours of
constant $\sin^2 2\theta_{13}$ shown in Fig.~\ref{fig:gut_plot}. This
implies a remarkable correlation between the predicted values of
$\Delta m^2_{21}$ and $\sin^2 2\theta_{13}$. This correlation is shown
explicitly in Fig. \ref{fig:theta13_dm21} which displays, for a grid of 
points that span the allowed region of the $(a,b)$-parameter space, the 
predicted values of
($\Delta m^2_{21}$, $\sin^2 2\theta_{13}$). The points are confined to a
narrow band, with $\sin^2 2\theta_{12}$ varying across the band. Note that
if the LMA solution is indeed the correct solution to explain the solar
neutrino deficit observations, KamLAND \cite{KamLAND} is expected to provide 
measurements
of $\Delta m^2_{21}$ and $\sin^2 2\theta_{12}$. Hence the GUT model we are
considering will be able to give a precise prediction for $\sin^2
2\theta_{13}$.

In Table~\ref{tab:x1} we have selected eight points in the LMA allowed
parameter region to illustrate the neutrino oscillation parameter predictions 
of the GUT model.  The correlations noted above are evident.

\begin{table}[t]
\caption[]{\label{tab:x1}
List of eight points selected in the LMA allowed parameter region to illustrate
the neutrino oscillation parameter predictions of the GUT model.\\}
\begin{tabular}{cccccccl}
  Point & \multicolumn{2}{c}{Model Parameters} & $\Delta m^2_{21}$ &
        $\Delta m^2_{32}$ &\hspace*{0.2in} $\sin^2 2\theta_{12}$&
        \hspace*{0.2in} $\sin^2 2\theta_{23}$ &\hspace*{0.2in}
        $\sin^2 2\theta_{13}$\\ 
        & $a$ & $b$ & ${\rm eV^2}$ & ${\rm eV^2}$ \\ \hline 
   (A)  & 1.0 & 2.0 & $6.5 \times 10^{-5}$ & $3.2 \times 10^{-3}$ 
        &\hspace*{0.2in} 0.880&\hspace*{0.2in} 0.994&\hspace*{0.2in} 0.0008 \\
   (B)  & 1.2 & 2.8 & $3.2 \times 10^{-5}$ & $3.2 \times 10^{-3}$ 
        &\hspace*{0.2in} 0.838&\hspace*{0.2in} 0.980&\hspace*{0.2in} 0.0038\\
   (C)  & 1.7 & 2.7 & $10.9 \times 10^{-5}$ & $3.2 \times 10^{-3}$ 
        &\hspace*{0.2in} 0.732&\hspace*{0.2in} 0.996&\hspace*{0.2in} 0.00008 \\
   (D)  & 1.7 & 3.0 & $6.3 \times 10^{-5}$ & $3.2 \times 10^{-3}$ 
        &\hspace*{0.2in} 0.745&\hspace*{0.2in} 0.999&\hspace*{0.2in} 0.0014 \\
   (E)  & 1.7 & 3.4 & $4.0 \times 10^{-5}$ & $3.2 \times 10^{-3}$ 
        &\hspace*{0.2in} 0.747&\hspace*{0.2in} 0.992&\hspace*{0.2in} 0.0033 \\
   (F)  & 2.0 & 3.0 & $12.8 \times 10^{-5}$ & $3.2 \times 10^{-5}$
        &\hspace*{0.2in} 0.655&\hspace*{0.2in} 0.987&\hspace*{0.2in} 0.00001 \\
   (G)  & 2.2 & 3.5 & $8.8 \times 10^{-5}$ & $3.2 \times 10^{-3}$ 
        &\hspace*{0.2in} 0.629&\hspace*{0.2in} 0.996&\hspace*{0.2in} 0.0008 \\
   (H)  & 2.2 & 4.3 & $3.6 \times 10^{-5}$ & $3.2 \times 10^{-3}$ 
        &\hspace*{0.2in} 0.648&\hspace*{0.2in} 0.993&\hspace*{0.2in} 0.0042 \\
  \end{tabular} 
\end{table}

We next consider the sensitivity of the predicted oscillation parameters 
to the assumed values of the underlying GUT model parameters. 
For a grid of points in the $(a,b)$-plane, Table~\ref{tab:x2} lists the 
$(\Delta a/a) / (\Delta \sin^2 2\theta_{12} / \sin^2 2\theta_{12})$, 
i.e., the fractional changes in the GUT scale parameter $a$ divided by 
the fractional changes in the predicted oscillation parameter 
$\sin^2 2 \theta_{12}$. The values vary from -1.2 to -5.5 
over the viable region of the $(a,b)$-plane. Hence, if the parameter $a$ 
is increased by 1\%, say, then the predicted value of $\sin^2 2\theta_{12}$ 
will typically decrease by a few percent. 
The corresponding sensitivity of the predicted value of $\Delta m^2_{21}$ 
to changes in $a$ is shown in Table~\ref{tab:x3}. Note that if the parameter 
$a$ is increased by 1\%, say, then the predicted value of $\Delta m^2_{21}$ 
increases typically by a fraction of a percent. Similar sensitivities are 
expected for the predicted values of $\sin^2 2\theta_{13}$ with changes 
in $a$ (Table~\ref{tab:x4}), or for the predicted values of 
$\Delta m^2_{21}$ (Table~\ref{tab:x5}) or $\sin^2 2\theta_{13}$ 
(Table~\ref{tab:x6}) with changes in $b$. 
The predicted values of $\sin^2 2\theta_{12}$ are insensitive to the value 
of $b$ (not shown in the tables). From these considerations we see that 
a precise measurement of $\sin^2 2\theta_{12}$ by KamLAND will precisely 
determine the GUT model parameter $a$ (for real $a$).  A very precise 
measurement of either $\Delta m^2_{21}$ 
or $\sin^2 2\theta_{13}$ will then precisely determine $b$.

In summary, our examination of the simplest case ($a$ and $b=c$ real) 
has revealed some striking features:
\begin{description}
\item{(i)} A large value for $\sin^2 2\theta_{13}$ cannot be accommodated.
In fact the model predicts $\sin^ 2\theta_{13} < 0.01$.
\item{(ii)} The prediction for $\sin^2 2\theta_{13}$ is precise once
$\Delta m^2_{21}$ and $\sin^2 2\theta_{12}$ are known.
\end{description}

\begin{table}[p]
\caption[]{\label{tab:x2}
Fractional change in the GUT scale parameter divided by the resulting 
fractional change in the oscillation parameter:  
$(\Delta a/a) / (\Delta \sin^2 2\theta_{12} / \sin^2 2\theta_{12})$.\\
}
\begin{tabular}{c|cccccc}
 & & & & a & & \\
b   &  1.2 &  1.4 &  1.6 &  1.8 &  2.0 &  2.2\\
\hline
4.5 &      &      &      &      &      & -1.7 \\
4.0 &      &      & -3.1 & -2.7 & -2.0 & -1.3 \\
3.5 &      & -4.6 & -3.2 & -2.0 & -1.4 &      \\
3.0 & -5.5 & -3.8 & -2.4 & -1.3 &      &      \\
2.5 & -4.0 & -2.6 & -1.2 &      &      &      \\
2.0 & -2.8 &      &      &      &      &      \\
\end{tabular}
\end{table}
%
%\newpage
\begin{table}[p]
\caption[]{\label{tab:x3}
Fractional change in the GUT scale parameter divided by the resulting
fractional change in the oscillation parameter:  
$(\Delta a/a) / (\Delta(\Delta m^2_{21}) / \Delta m^2_{21})$.\\
}
\begin{tabular}{c|cccccc}
 & & & & a & & \\
b   &  1.2 &  1.4 &  1.6 &  1.8 &  2.0 &  2.2\\
\hline
4.5 &     &     &     &      &      & 0.3  \\
4.0 &     &     & 0.4 & 0.4  & 0.3  & 0.3  \\
3.5 &     & 0.5 & 0.4 & 0.4  & 0.3  &      \\
3.0 & 0.5 & 0.4 & 0.3 & 0.2  &      &      \\
2.5 & 0.3 & 0.2 &     &      &      &      \\
2.0 & 0.2 &     &     &      &      &      \\
\end{tabular}
\end{table}

%\newpage
\begin{table}[p]
\caption[]{\label{tab:x4}
Fractional change in the GUT scale parameter divided by the resulting
fractional change in the oscillation parameter:  
$(\Delta a/a) / (\Delta \sin^2 2\theta_{13} / \sin^2 2\theta_{13})$.\\
}
\begin{tabular}{c|cccccc}
\hline
 & & & & a & & \\
b   &  1.2 &  1.4 &  1.6 &  1.8 &  2.0 &  2.2\\
\hline
4.5 &      &      &      &      & -0.7 & -0.6 \\
4.0 &      &      &      & -0.7 & -0.4 & -0.3 \\
3.5 &      & -0.8 & -0.5 & -0.3 & -0.2 & -0.1 \\
3.0 & -0.6 & -0.4 & -0.2 & -0.1 &      &      \\
2.5 & -0.2 & -0.1 &      &      &      &      \\
2.0 &      &      &      &      &      &      \\
\end{tabular}
\end{table}
\newpage
\begin{table}[t]
\caption[]{\label{tab:x5}
Fractional change in the GUT scale parameter divided by the resulting
fractional change in the oscillation parameter:  
$(\Delta b/b) / (\Delta(\Delta m^2_{21}) / \Delta m^2_{21})$.\\ 
}
\begin{tabular}{c|cccccc}
 & & & & a & & \\
b   &  1.2 &  1.4 &  1.6 &  1.8 &  2.0 &  2.2\\
\hline
4.5 &      &      &      &      &      &      \\
4.0 &      &      &      &      &      & -0.2 \\
3.5 &      &      & -0.1 & -0.2 & -0.2 &      \\
3.0 & -0.2 & -0.2 & -0.2 &      &      &      \\
2.5 & -0.2 & -0.4 &      &      &      &      \\
2.0 & -0.4 &      &      &      &      &      \\
\end{tabular}
\end{table}
\begin{table}[t]
\caption[]{\label{tab:x6}
Fractional change in the GUT scale parameter divided by the resulting 
fractional change in the oscillation parameter:  
$(\Delta b/b) / (\Delta \sin^2 2\theta_{13} / \sin^2 2\theta_{13})$.\\
}
\begin{tabular}{c|cccccc}
 & & & & a & & \\
b   &  1.2 &  1.4 &  1.6 &  1.8 &  2.0 &  2.2\\
\hline
4.5 &     &      &     &       & 0.4  & 0.3  \\
4.0 &     &      &     & 0.4   & 0.3  & 0.2  \\
3.5 &     & 0.4  & 0.3 & 0.2   & 0.1  & 0.05 \\
3.0 & 0.3 & 0.3  & 0.2 & 0.05  &      &      \\
2.5 & 0.2 & 0.05 &     &       &      &      \\
2.0 &     &      &     &       &      &      \\
\end{tabular}
\end{table}

\subsection{Parameter Choice: $b = c$ Real with $a$ Complex}
\begin{figure}[h]
\centering\leavevmode
\epsfxsize=6.0in\epsffile{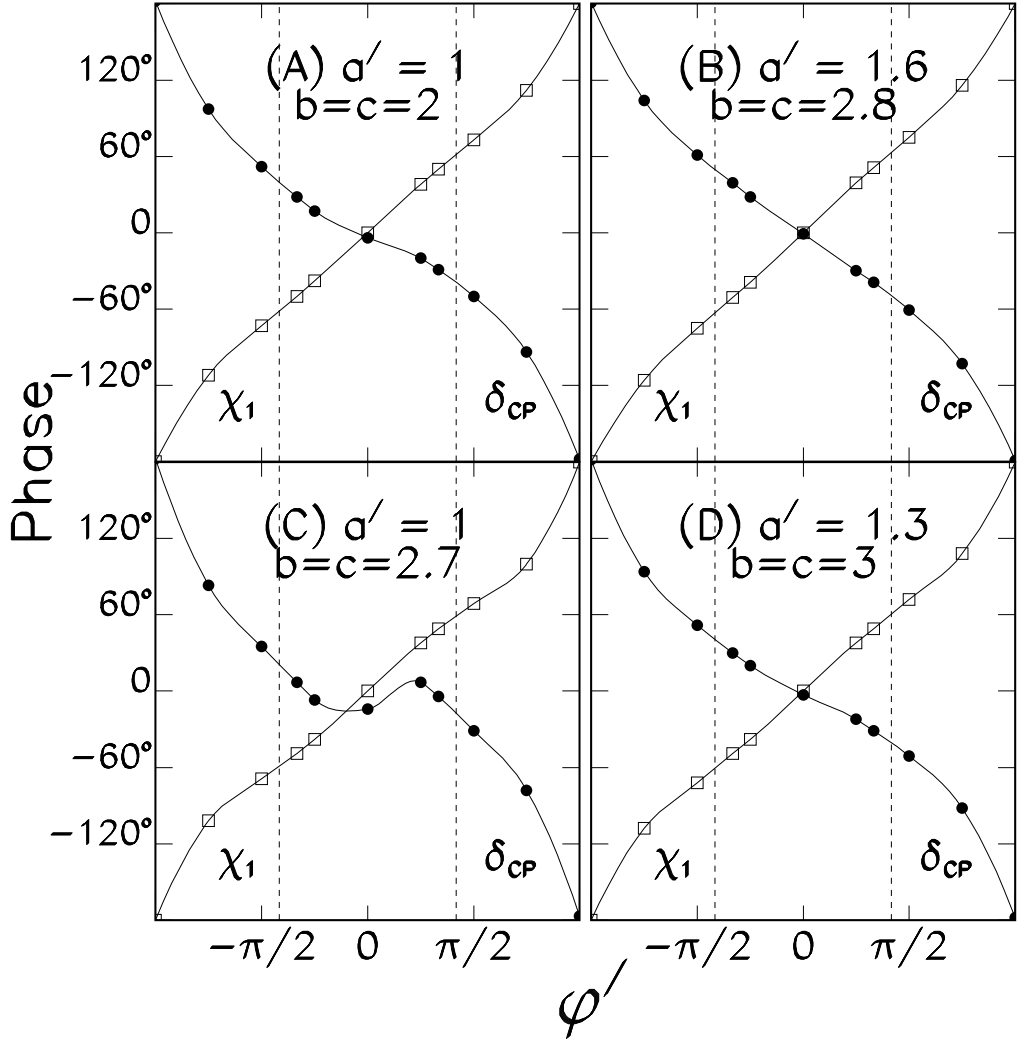}
\medskip
\vspace{0.5cm}
\caption[]{The observable CP phase $\delta_{CP}$ and Majorana phase 
$\chi_1$ are shown as functions of the
GUT phase parameter $\phi'$ for the first four of the eight points in parameter
space that are listed in Table~\ref{tab:x1}.  The ranges of $\phi'$ of interest
lie between the dashed lines.}
\label{fig:delta_phi_1}
\end{figure}

\begin{figure}[h]
\centering\leavevmode
\epsfxsize=6.0in\epsffile{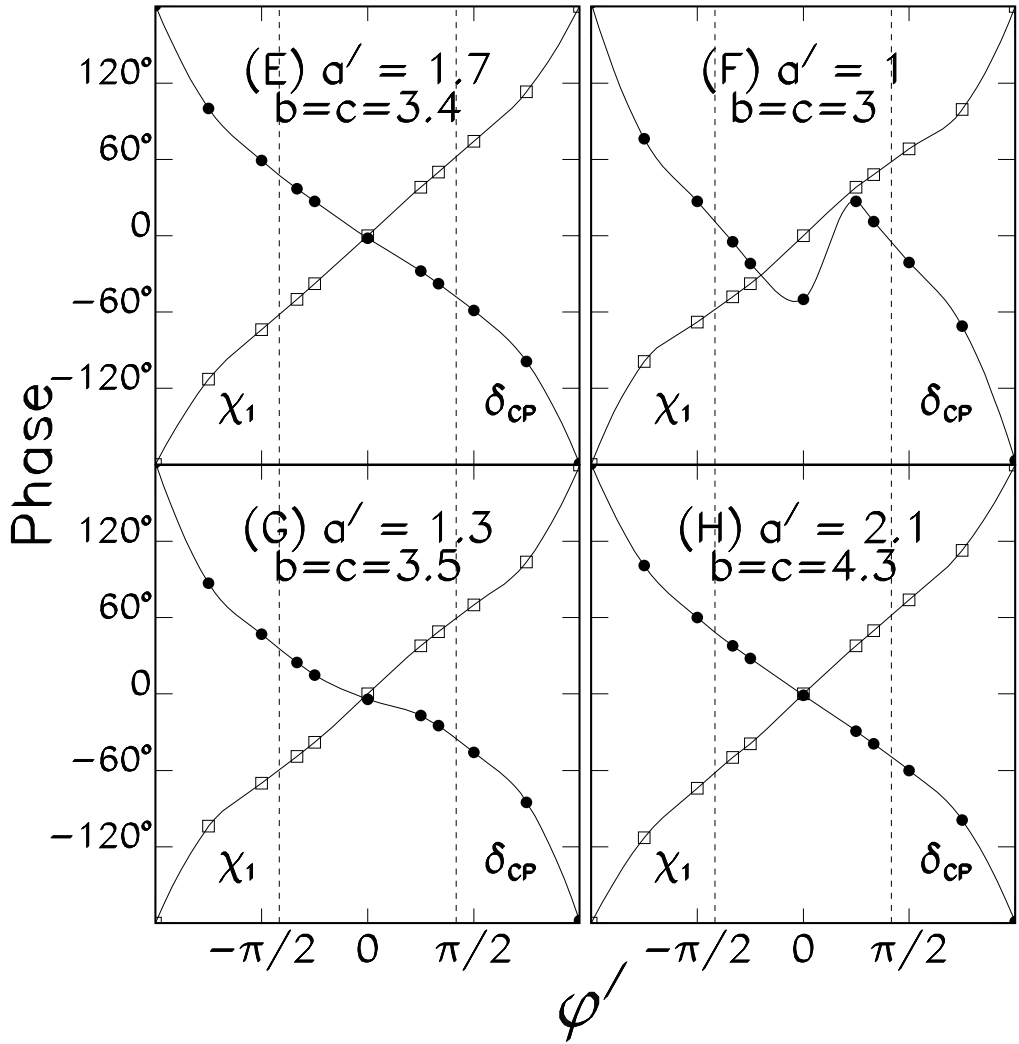}
\medskip
\vspace{0.5cm}
\caption[]{The observable CP phase $\delta_{CP}$ and Majorana phase 
$\chi_1$ are shown as functions of the
GUT phase parameter $\phi'$ for the second four of the eight points in parameter
space that are listed in Table~\ref{tab:x1}.  The ranges of $\phi'$ of interest
lie between the dashed lines.}
\label{fig:delta_phi_2}
\end{figure}

We have seen from the example presented in Sec. III that the CP phase,
$\delta_{CP}$, turns out to be very small, since both the Dirac neutrino
matrix $N$ and the right-handed Majorana matrix $M_R$ are real,
while only the charged lepton matrix $L$ is complex and results in a small
complex contribution to $U_{MNS}$.  But with two Higgs lepton-violating 
singlets contributing to $M_R$, one can 
introduce an additional complex phase $\phi'$ into $M_R$.  In discussing CP 
violation, we shall identify
\begin{equation}
	a \equiv b - a'e^{i\phi'},\ {\rm with}\ b = c,\\
\label{eq:cpparam}
\end{equation}
where $b$ is real and arises from the first Higgs singlet which contributes 
to all nine matrix elements of $M_R$, while $a'$ can be complex and arises
from the second Higgs singlet which contributes to only the 13 and 31 
elements.  Any observable CP violation in the lepton sector with 
its phase $\delta_{CP}$ is then controlled by $\phi'$ and the phase $\phi$ 
appearing in the charged lepton matrix $L$ in Eq. (\ref{eq:Dirac}).
The viable region of parameter space shown in Fig.~\ref{fig:gut_plot} and
\ref{fig:gut_plot2} is not significantly changed.  To understand
the predictions in detail, we again choose the eight specific
points in parameter space listed earlier in Table~\ref{tab:x1}.
For each point, the
predictions for $\sin^2 2\theta_{12}$, $\sin^2 2\theta_{23}$,
$\sin^2 2\theta_{13}$, and $\delta_{CP}$ are listed as functions of
$\phi'$ in Tables VII - XIV.  The predicted observable $\delta_{CP}$, as
well as the Majorana phase $\chi_1$, are
shown for each point as functions of $\phi'$ in 
Figs.~\ref{fig:delta_phi_1} and~\ref{fig:delta_phi_2}.  Only the range
$|\phi'| < 75^\circ$ within the dashed lines in these figures is consistent
with the present lower limit on $\sin^2 2\theta_{23}$. 
%
%\newpage
\noindent Note that when $\phi' = 0$, corresponding to the maximum values 
predicted for $\sin^2 2\theta_{23}$, the predictions for $\delta_{CP}$ are 
typically a few degrees except for cases (C) and (F) for which 
$\delta_{CP} = -14^\circ$ and $-50^\circ$, respectively.  The peculiar behavior
for these two special cases arises because $\sin^2 2\theta_{23}$ becomes 
maximal and crosses from the dark side ($\tan \theta_{23} > 1$) into the light 
side ($\tan \theta_{23} < 1$) and back into the dark side as $\phi'$ goes 
through $0^\circ$.  The predictions for $\chi_1$ and $\chi_2$, on the other 
hand, 
are smoothly varying in all cases.  Since $\Delta m^2_{21}$ is on the high 
side of the allowed region for cases (C) and (F), and somewhat disfavored by 
other recent analyses \cite{LMA}, it appears that the GUT model under 
consideration predicts that leptonic CP violation will be small
for the near maximal values of $\sin^2 2\theta_{23}$ and more generally that
$|\delta_{CP}| < 50^\circ$.

\begin{figure}[h]
\centering\leavevmode
\epsfxsize=5.5in\epsffile{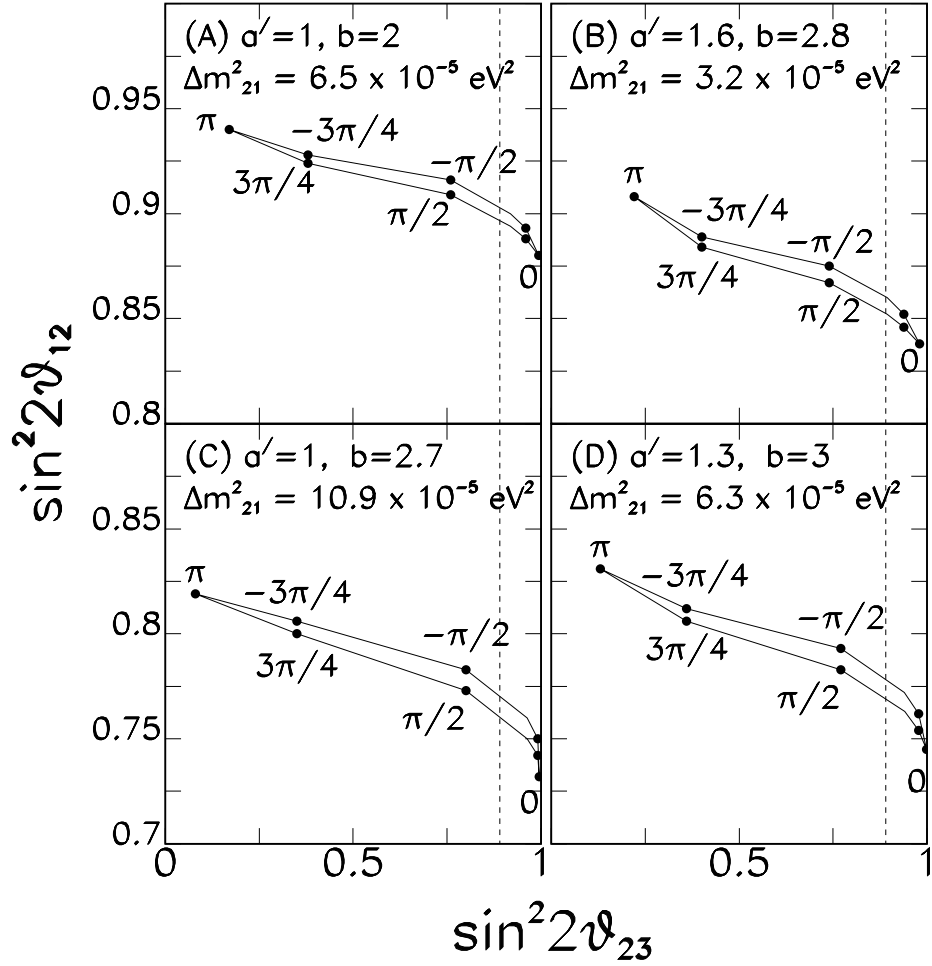}
\medskip
\vspace{0.5cm}
\caption[]{The predicted value of $\sin^2 2\theta_{12}$ shown as
a function of the predicted $\sin^2 2\theta_{23}$ for the first four of
the eight points in parameter space that are listed in Table~\ref{tab:x1}.
The values of $\delta_{CP}$ vary around the contour of solutions and are
indicated at points corresponding to $\phi' = 0,\ \pm\pi/4,\ \pm\pi/2, 
\ \pm 3\pi/4$, and $\pi$.  The viable region in $\sin^2 2\theta_{23}$ lies 
between 0.89 and 1.0.}
\label{fig:cp_contour_1}
\end{figure}

\begin{figure}[p]
\centering\leavevmode
\epsfxsize=5.5in\epsffile{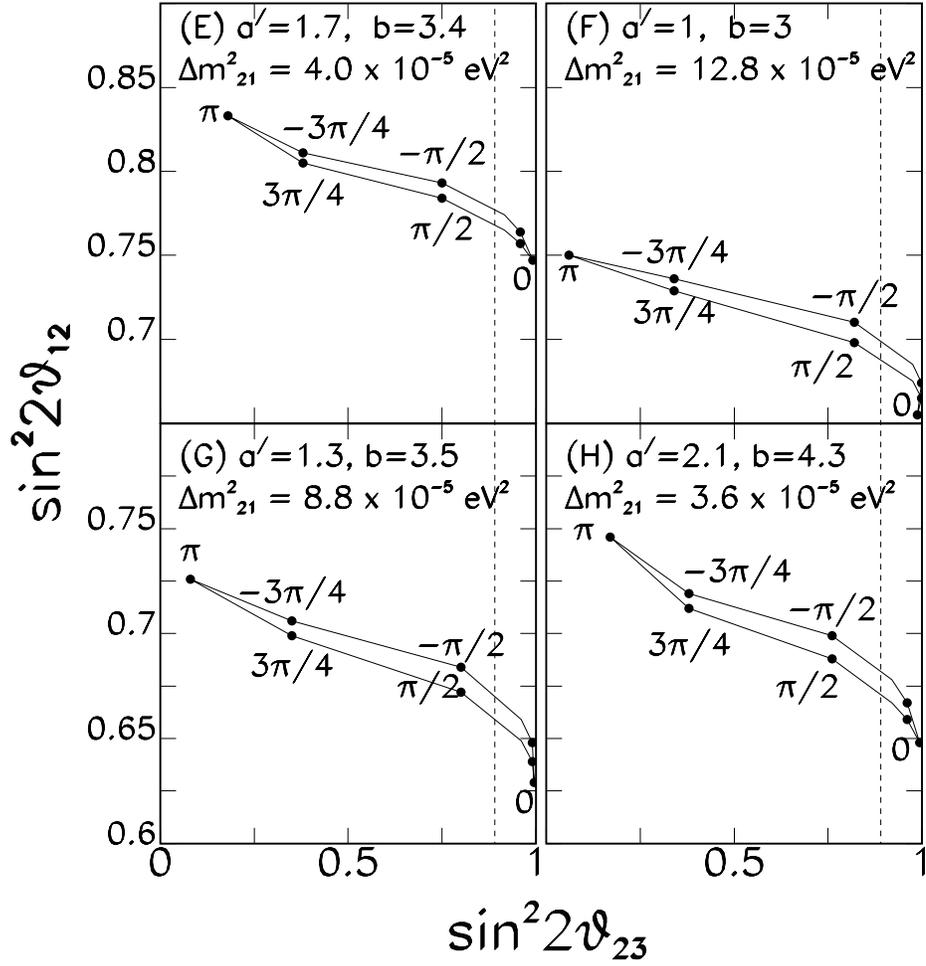}
\medskip
\vspace{0.5cm}
\caption[]{The predicted value of $\sin^2 2\theta_{12}$ shown as
a function of the predicted $\sin^2 2\theta_{23}$ for the second four of the
eight points in parameter space that are listed in Table~\ref{tab:x1}.
The values of $\delta_{CP}$ vary around the contour of solutions and are
indicated at points corresponding to $\phi' = 0,\ \pm\pi/4,\ \pm\pi/2, 
\ \pm 3\pi/4$, and $\pi$.}
\label{fig:cp_contour_2}
\end{figure}

Next consider the predictions for the mixing angles, $\sin^2 2\theta_{12}$ 
and $\sin^2 2\theta_{23}$, which are shown for the 8 points in parameter 
space in Figs.~\ref{fig:cp_contour_1} and~\ref{fig:cp_contour_2}.
These figures show the predictions as functions of $\phi'$.
Within the viable region of parameter space corresponding to
$\sin^2 2\theta_{23} > 0.89$, the permitted values of
$\sin^2 2\theta_{12}$ are restricted for each point in
$(a^\prime, b)$-space. A 10\% measurement of $\sin^2 2\theta_{12}$ by the 
KamLAND experiment, combined with a few percent
measurement of $\sin^2 2\theta_{23}$ by MINOS and the CNGS
experiments would enable significant 
regions of the GUT model parameter space to be excluded. A 1\% 
measurement of $\sin^2 2\theta_{23}$ at a Neutrino Factory would provide a
stringent test of the GUT model.

\subsection{Parameter Choice: $a \neq b \neq c$}

The more general GUT model case with $a \neq b \neq c$ would arise if three
Higgs VEVs breaking lepton number were to contribute to the right-handed
Majorana mass matrix.  This complication is much more difficult to analyse
and is not studied here.  The two simplified cases we have studied appear 
sufficient to present a realistic picture of neutrino oscillations.

\section{Conclusions}

Within the framework of an $SO(10)$ GUT model developed by Albright and Barr
that can accommodate both the atmospheric and LMA solar neutrino mixing
solutions, we have presented explicit predictions for $\sin^2 2\theta_{13}$, 
$\sin^2 2\theta_{12}$, $\sin^2 2\theta_{23}$, and $\Delta m^2_{21}$.
Precise measurements of $\sin^2 2\theta_{12}$ and $\Delta m^2_{21}$ by 
KamLAND can be used to precisely determine the GUT parameters $a$ 
(with $a$ real) and $b$.  We find that the model can then be tested with 
precision neutrino oscillation measurements of $\sin^2 2\theta_{23}$, 
$\sin^2 2\theta_{13}$, and the leptonic CP phase $\delta^\prime$ at 
Neutrino Superbeams and Neutrino Factories. 

Over the entire region of viable GUT model 
parameter space, the value of $\sin^2 2\theta_{13}$ 
is predicted to be less than 0.01. If this is the 
case, $\nu_\mu \to \nu_e$ oscillations will not be 
observed by the MINOS or CNGS experiments. 
Over half of the viable parameter space, the predicted 
$\sin^2 2\theta_{13}$ exceeds 0.003, and 
$\nu_\mu \to \nu_e$ oscillations would be expected to 
be observed at Neutrino Superbeams. The remaining half 
of the parameter space would be probed at a Neutrino Factory, 
except a small region for which $\sin^2 2\theta_{13} < 0.0001$. 
The GUT model predicts a striking correlation 
between $\Delta m^2_{21}$ and $\sin^2 2\theta_{13}$. 
Once $\Delta m^2_{21}$ is measured by 
KamLAND with a precision of a few percent, the 
model will predict $\sin^2 2\theta_{13}$ with 
a precision of a few percent. A precise test 
of the model with this level of precision will 
require a Neutrino Factory.

In the more general version of the GUT model in which 
$a$ is complex, the absolute observable CP phase $|\delta_{CP}|$ 
is at most $\sim 50^\circ$ over almost the entire viable 
parameter space.  For the maximal atmospheric neutrino mixing region,
$\delta_{CP}$ is typically very small, with exceptions noted earlier for 
the largest values of $\Delta m^2_{21}$ presently allowed.  
The predicted $\langle m_{\beta\beta} \rangle$ is at most a few times
$10^{-4}$ eV, too small for neutrinoless double beta decay to be observed
by the next generation experiments.

Finally, a general conclusion from the study of the 
predictions of one specific GUT model is that, if the 
LMA solar solution is confirmed, very 
precise measurements of all the oscillation parameters 
are important to test the theory and determine the 
associated parameters. We will need a Neutrino Factory.\\[0.2in]

The initial preparation of this manuscript was carried out at the Snowmass 2001
Workshop on the Future of Particle Physics.  One of us (CHA) thanks 
Stephen Barr for several discussions on the complex extension of the 
right-handed Majorana neutrino mass matrix that was developed in 
collaboration with him for the LMA solution.

\newpage
\begin{table}\label{tab:x7}
\caption[]{Oscillation Parameters for Point (A)\\}
\begin{tabular}{rcllrrr}
	\multicolumn{7}{l}{{\bf (A) $M_R$ Model Parameters:} \quad $a' = 1.0,
		\ b = c = 
		2.0,\ \Lambda_R = 2.5 \times 10^{14}$ GeV}\\[0.1in]
	\multicolumn{7}{c}{\hspace*{0.3in}{\bf Predictions:}\quad 
		$\Delta m^2_{21} = 6.5\times 10^{-5}\ {\rm eV^2},\quad 
		\Delta m^2_{32} = 3.2 \times 10^{-3}\ {\rm eV^2}$}\\[0.1in]
	$\phi'$\hspace*{0.1in} &\hspace*{0.4in} $\sin^2 2\theta_{12}$ 
		&\hspace*{0.4in} $\sin^2 2\theta_{23}$ & \hspace*{0.4in} 
		$\sin^2 2\theta_{13}$ & $\delta_{CP}$\hspace*{0.5in} & 
		$\chi_{1}$ \hspace*{0.5in} & $\chi_{2}$\hspace*{0.10in}\\
		\hline\hline
	$-\pi$ &\hspace*{0.4in} 0.940 &\hspace*{0.5in} 0.17
		&\hspace*{0.5in} 0.0033 & \hspace*{0.4in} $182^\circ$ 
		\hspace*{0.4in} & $-180^\circ$ \hspace*{0.4in} 
		& $-180^\circ$ \\
	$-3\pi/4$ &\hspace*{0.4in} 0.928 &\hspace*{0.5in} 0.38
		&\hspace*{0.5in} 0.0031 & \hspace*{0.4in} $97^\circ$ 
		\hspace*{0.4in} & $-112^\circ$ \hspace*{0.4in}
		& $-109^\circ$\\
	$-\pi/2$ &\hspace*{0.4in} 0.916 &\hspace*{0.5in} 0.76
		&\hspace*{0.5in} 0.0022 & \hspace*{0.4in} $52^\circ$ 
		\hspace*{0.4in} & $-73^\circ$ \hspace*{0.4in} 
		& $-69^\circ$ \\
	\hline
	$-\pi/3$ &\hspace*{0.4in} 0.900 &\hspace*{0.5in} 0.919
		&\hspace*{0.5in} 0.0015 & \hspace*{0.4in} $28^\circ$ 
		\hspace*{0.4in} & $-50^\circ$ \hspace*{0.4in} 
		& $-47^\circ$\\ 
	$-\pi/4$ &\hspace*{0.4in} 0.893 &\hspace*{0.5in} 0.960
		&\hspace*{0.5in} 0.0013 & \hspace*{0.4in} $17^\circ$ 
		\hspace*{0.4in} & $-38^\circ$ \hspace*{0.4in} 
		& $-35^\circ$ \\
	$0$ &\hspace*{0.4in} 0.880 &\hspace*{0.5in} 0.994
		&\hspace*{0.5in} 0.0008 & \hspace*{0.4in} $-4^\circ$ 
		\hspace*{0.4in} & $0^\circ$ \hspace*{0.4in} 
		& $0^\circ$ \\
	$\pi/4$ &\hspace*{0.4in} 0.888 &\hspace*{0.5in} 0.960
		&\hspace*{0.5in} 0.0010 & \hspace*{0.4in} $-20^\circ$ 
		\hspace*{0.4in} & $38^\circ$ \hspace*{0.4in} 
		& $35^\circ$\\
	$\pi/3$ &\hspace*{0.4in} 0.894 &\hspace*{0.5in} 0.919
		&\hspace*{0.5in} 0.0013 & \hspace*{0.4in} $-29^\circ$ 
		\hspace*{0.4in} & $50^\circ$ \hspace*{0.4in} 
		& $47^\circ$\\
	\hline	
	$\pi/2$ &\hspace*{0.4in} 0.909 &\hspace*{0.5in} 0.76
		&\hspace*{0.5in} 0.0019 & \hspace*{0.4in} $-50^\circ$ 
		\hspace*{0.4in} & $73^\circ$ \hspace*{0.4in} 
		& $69^\circ$\\
	$3\pi/4$ &\hspace*{0.4in} 0.924 &\hspace*{0.5in} 0.38
		&\hspace*{0.5in} 0.0028 & \hspace*{0.4in} $-94^\circ$ 
		\hspace*{0.4in} & $112^\circ$ \hspace*{0.4in} 
		& $109^\circ$ \\
	$\pi$ &\hspace*{0.4in} 0.940 &\hspace*{0.5in} 0.17
		&\hspace*{0.5in} 0.0033 & \hspace*{0.4in} $-178^\circ$ 
		\hspace*{0.4in} & $180^\circ$ \hspace*{0.4in} 
		& $180^\circ$\\
\end{tabular}
\end{table}

\begin{table}\label{tab:x8}
\caption[]{Oscillation Parameters for Point (B)\\}
\begin{tabular}{rcllrrr}
	\multicolumn{7}{l}{{\bf (B) $M_R$ Model Parameters:} \quad $a' = 1.6,
		\ b = c = 2.8,\ \Lambda_R = 2.4 \times 10^{14}$ GeV}\\[0.1in]
	\multicolumn{7}{c}{\hspace*{0.3in}{\bf Predictions:}\quad 
		$\Delta m^2_{21} = 3.2\times 10^{-5}\ {\rm eV^2},\quad 
		\Delta m^2_{32} = 3.2 \times 10^{-3}\ {\rm eV^2}$}\\[0.1in]
	$\phi'$\hspace*{0.1in} &\hspace*{0.4in} $\sin^2 2\theta_{12}$ 
		&\hspace*{0.4in} $\sin^2 2\theta_{23}$ & \hspace*{0.4in} 
		$\sin^2 2\theta_{13}$ & $\delta_{CP}$\hspace*{0.5in} & 
		$\chi_{1}$ \hspace*{0.5in} & $\chi_{2}$\hspace*{0.10in}\\
		\hline\hline
	$-\pi$ &\hspace*{0.4in} 0.908 &\hspace*{0.5in} 0.22
		&\hspace*{0.5in} 0.0077 & \hspace*{0.4in} $181^\circ$ 
		\hspace*{0.4in} & $-180^\circ$ \hspace*{0.4in} 
		& $-180^\circ$\\
	$-3\pi/4$ &\hspace*{0.4in} 0.889 &\hspace*{0.5in} 0.40
		&\hspace*{0.5in} 0.0073 & \hspace*{0.4in} $104^\circ$ 
		\hspace*{0.4in} & $-116^\circ$ \hspace*{0.4in} 
		& $-114^\circ$ \\
	$-\pi/2$ &\hspace*{0.4in} 0.875 &\hspace*{0.5in} 0.74
		&\hspace*{0.5in} 0.0060 & \hspace*{0.4in} $61^\circ$ 
		\hspace*{0.4in} & $-75^\circ$ \hspace*{0.4in} 
		& $-72^\circ$\\
	\hline
	$-\pi/3$ &\hspace*{0.4in} 0.860 &\hspace*{0.5in} 0.894
		&\hspace*{0.5in} 0.0049 & \hspace*{0.4in} $39^\circ$ 
		\hspace*{0.4in} & $-51^\circ$ \hspace*{0.4in} 
		& $-48^\circ$ \\ 
	$-\pi/4$ &\hspace*{0.4in} 0.852 &\hspace*{0.5in} 0.938
		&\hspace*{0.5in} 0.0045 & \hspace*{0.4in} $28^\circ$ 
		\hspace*{0.4in} & $-39^\circ$ \hspace*{0.4in} 
		& $-36^\circ$\\
	$0$ &\hspace*{0.4in} 0.838 &\hspace*{0.5in} 0.980
		&\hspace*{0.5in} 0.0038 & \hspace*{0.4in} $-1^\circ$ 
		\hspace*{0.4in} & $0^\circ$ \hspace*{0.4in} 
		& $0^\circ$\\
	$\pi/4$ &\hspace*{0.4in} 0.846 &\hspace*{0.5in} 0.938
		&\hspace*{0.5in} 0.0042 & \hspace*{0.4in} $-30^\circ$ 
		\hspace*{0.4in} & $39^\circ$ \hspace*{0.4in} 
		& $36^\circ$\\
	$\pi/3$ &\hspace*{0.4in} 0.852 &\hspace*{0.5in} 0.894
		&\hspace*{0.5in} 0.0045 & \hspace*{0.4in} $-39^\circ$ 
		\hspace*{0.4in} & $51^\circ$ \hspace*{0.4in} 
		& $48^\circ$ \\
	\hline
	$\pi/2$ &\hspace*{0.4in} 0.867 &\hspace*{0.5in} 0.74
		&\hspace*{0.5in} 0.0055 & \hspace*{0.4in} $-61^\circ$ 
		\hspace*{0.4in} & $75^\circ$ \hspace*{0.4in} 
		& $72^\circ$ \\
	$3\pi/4$ &\hspace*{0.4in} 0.884 &\hspace*{0.5in} 0.40
		&\hspace*{0.5in} 0.0069 & \hspace*{0.4in} $-103^\circ$ 
		\hspace*{0.4in} & $116^\circ$ \hspace*{0.4in} 
		& $114^\circ$ \\
	$\pi$ &\hspace*{0.4in} 0.908 &\hspace*{0.5in} 0.22
		&\hspace*{0.5in} 0.0077 & \hspace*{0.4in} $-179^\circ$ 
		\hspace*{0.4in} & $180^\circ$ \hspace*{0.4in} 
		& $180^\circ$\\
\end{tabular}
\end{table}	

\begin{table}\label{tab:x9}
\caption[]{Oscillation Parameters for Point (C)\\}
\begin{tabular}{rcllrrr}
	\multicolumn{7}{l}{{\bf (C) $M_R$ Model Parameters:} \quad $a' = 1.0,
		\ b = c = 2.7,\ \Lambda_R = 2.5 \times 10^{14}$ GeV}\\[0.1in]
	\multicolumn{7}{c}{\hspace*{0.3in}{\bf Predictions:}\quad 
		$\Delta m^2_{21} = 10.9 \times 10^{-5}\ {\rm eV^2},\quad 
		\Delta m^2_{32} = 3.2 \times 10^{-3}\ {\rm eV^2}$}\\[0.1in]
	$\phi'$\hspace*{0.1in} &\hspace*{0.4in} $\sin^2 2\theta_{12}$ 
		&\hspace*{0.4in} $\sin^2 2\theta_{23}$ & \hspace*{0.4in} 
		$\sin^2 2\theta_{13}$ & $\delta_{CP}$\hspace*{0.5in} & 
		$\chi_{1}$ \hspace*{0.5in} & $\chi_{2}$\hspace*{0.10in}\\
		\hline\hline
	$-\pi$ &\hspace*{0.4in} 0.819 &\hspace*{0.5in} 0.08 
		&\hspace*{0.5in} 0.0021 & \hspace*{0.4in} $183^\circ$
		\hspace*{0.4in} & $-180^\circ$ \hspace*{0.4in}
		& $-180^\circ$\\
	$-3\pi/4$ &\hspace*{0.4in} 0.806 &\hspace*{0.5in} 0.35 
		&\hspace*{0.5in} 0.0019 & \hspace*{0.4in} $83^\circ$ 
		\hspace*{0.4in} & $-102^\circ$ \hspace*{0.4in} 
		& $-100^\circ$
		 \\
	$-\pi/2$ &\hspace*{0.4in} 0.783 &\hspace*{0.5in} 0.80 
		&\hspace*{0.5in} 0.0012 & \hspace*{0.4in} $35^\circ$ 
		\hspace*{0.4in} & $-69^\circ$ \hspace*{0.4in} 
		& $-66^\circ$\\
	\hline
	$-\pi/3$ &\hspace*{0.4in} 0.760 &\hspace*{0.5in} 0.963 
		&\hspace*{0.5in} 0.0007 & \hspace*{0.4in} $7^\circ$ 
		\hspace*{0.4in} & $-49^\circ$ \hspace*{0.4in} 
		& $-45^\circ$\\ 
	$-\pi/4$ &\hspace*{0.4in} 0.750 &\hspace*{0.5in} 0.992 
		&\hspace*{0.5in} 0.0005 & \hspace*{0.4in} $-7^\circ$ 
		\hspace*{0.4in} & $-38^\circ$ \hspace*{0.4in} 
		& $-35^\circ$ \\
	$0$ &\hspace*{0.4in} 0.732 &\hspace*{0.5in} 0.996 
		&\hspace*{0.5in} 0.0001 & \hspace*{0.4in} $-14^\circ$ 
		\hspace*{0.4in} & $0^\circ$ \hspace*{0.4in} 
		& $0^\circ$\\
	$\pi/4$ &\hspace*{0.4in} 0.742 &\hspace*{0.5in} 0.992 
		&\hspace*{0.5in} 0.0003 & \hspace*{0.4in} $7^\circ$ 
		\hspace*{0.4in} & $38^\circ$ \hspace*{0.4in} 
		& $35^\circ$\\
	$\pi/3$ &\hspace*{0.4in} 0.750 &\hspace*{0.5in} 0.963 
		&\hspace*{0.5in} 0.0005 & \hspace*{0.4in} $-4^\circ$ 
		\hspace*{0.4in} & $49^\circ$ \hspace*{0.4in} & $45^\circ$\\
	\hline
	$\pi/2$ &\hspace*{0.4in} 0.773 &\hspace*{0.5in} 0.80 
		&\hspace*{0.5in} 0.0010 & \hspace*{0.4in} $-31^\circ$ 
		\hspace*{0.4in} & $69^\circ$ \hspace*{0.4in} & $66^\circ$\\
	$3\pi/4$ &\hspace*{0.4in} 0.800 &\hspace*{0.5in} 0.35 
		&\hspace*{0.5in} 0.0017 & \hspace*{0.4in} $-78^\circ$ 
		\hspace*{0.4in} & $100^\circ$ \hspace*{0.4in} & $102^\circ$\\
	$\pi$ &\hspace*{0.4in} 0.819 &\hspace*{0.5in} 0.08 
		&\hspace*{0.5in} 0.0021 & \hspace*{0.4in} $-177^\circ$ 
		\hspace*{0.4in} & $180^\circ$ \hspace*{0.4in} & $180^\circ$\\
\end{tabular}
\end{table}

\begin{table}\label{tab:x10}
\caption[]{Oscillation Parameters for Point (D)\\}
\begin{tabular}{rcllrrr}
	\multicolumn{7}{l}{{\bf (D) $M_R$ Model Parameters:} \quad $a' = 1.3,
		\ b = c = 
		3.0,\ \Lambda_R = 2.5 \times 10^{14}$ GeV}\\[0.1in]
	\multicolumn{7}{c}{\hspace*{0.3in}{\bf Predictions:}\quad 
		$\Delta m^2_{21} = 6.3\times 10^{-5}\ {\rm eV^2},\quad 
		\Delta m^2_{32} = 3.2 \times 10^{-3}\ {\rm eV^2}$}\\[0.1in]
	$\phi'$\hspace*{0.1in} &\hspace*{0.4in} $\sin^2 2\theta_{12}$ 
		&\hspace*{0.4in} $\sin^2 2\theta_{23}$ & \hspace*{0.4in} 
		$\sin^2 2\theta_{13}$ & $\delta_{CP}$\hspace*{0.5in} & 
		$\chi_{1}$ \hspace*{0.5in} & $\chi_{2}$\hspace*{0.10in}\\
		\hline\hline
	$-\pi$ &\hspace*{0.4in} 0.831 &\hspace*{0.5in} 0.13
		&\hspace*{0.5in} 0.0049 & \hspace*{0.4in} $182^\circ$ 
		\hspace*{0.4in} & $-180^\circ$ \hspace*{0.4in} 
		& $-180^\circ$\\
	$-3\pi/4$ &\hspace*{0.4in} 0.812 &\hspace*{0.5in} 0.36
		&\hspace*{0.5in} 0.0045 & \hspace*{0.4in} $94^\circ$ 
		\hspace*{0.4in} & $-108^\circ$ \hspace*{0.4in} 
		& $-106^\circ$\\
	$-\pi/2$ &\hspace*{0.4in} 0.793 &\hspace*{0.5in} 0.77
		&\hspace*{0.5in} 0.0033 & \hspace*{0.4in} $52^\circ$ 
		\hspace*{0.4in} & $-72^\circ$ \hspace*{0.4in} 
		& $-68^\circ$\\
	\hline
	$-\pi/3$ &\hspace*{0.4in} 0.772 &\hspace*{0.5in} 0.939
		&\hspace*{0.5in} 0.0024 & \hspace*{0.4in} $30^\circ$ 
		\hspace*{0.4in} & $-49^\circ$ \hspace*{0.4in} 
		& $-46^\circ$\\ 
	$-\pi/4$ &\hspace*{0.4in} 0.762 &\hspace*{0.5in} 0.977
		&\hspace*{0.5in} 0.0020 & \hspace*{0.4in} $20^\circ$ 
		\hspace*{0.4in} & $-38^\circ$ \hspace*{0.4in} 
		& $-35^\circ$\\
	$0$ &\hspace*{0.4in} 0.745 &\hspace*{0.5in} 0.9991
		&\hspace*{0.5in} 0.0014 & \hspace*{0.4in} $-3^\circ$ 
		\hspace*{0.4in} & $0^\circ$ \hspace*{0.4in} 
		& $0^\circ$\\
	$\pi/4$ &\hspace*{0.4in} 0.754 &\hspace*{0.5in} 0.977
		&\hspace*{0.5in} 0.0017 & \hspace*{0.4in} $-22^\circ$ 
		\hspace*{0.4in} & $38^\circ$ \hspace*{0.4in} 
		& $35^\circ$\\
	$\pi/3$ &\hspace*{0.4in} 0.763 &\hspace*{0.5in} 0.940
		&\hspace*{0.5in} 0.0021 & \hspace*{0.4in} $-31^\circ$ 
		\hspace*{0.4in} & $49^\circ$ \hspace*{0.4in} 
		& $46^\circ$ \\
	\hline
	$\pi/2$ &\hspace*{0.4in} 0.783 &\hspace*{0.5in} 0.77
		&\hspace*{0.5in} 0.0029 & \hspace*{0.4in} $-51^\circ$ 
		\hspace*{0.4in} & $72^\circ$ \hspace*{0.4in} 
		& $68^\circ$\\
	$3\pi/4$ &\hspace*{0.4in} 0.806 &\hspace*{0.5in} 0.36
		&\hspace*{0.5in} 0.0042 & \hspace*{0.4in} $-92^\circ$ 
		\hspace*{0.4in} & $108^\circ$ \hspace*{0.4in} 
		& $106^\circ$\\
	$\pi$ &\hspace*{0.4in} 0.831 &\hspace*{0.5in} 0.13
		&\hspace*{0.5in} 0.0049 & \hspace*{0.4in} $-178^\circ$ 
		\hspace*{0.4in} & $180^\circ$ \hspace*{0.4in} 
		& $180^\circ$ \\
\end{tabular}
\end{table}

\begin{table}\label{tab:x11}
\caption[]{Oscillation Parameters for Point (E)\\}
\begin{tabular}{rcllrrr}
	\multicolumn{7}{l}{{\bf (E) $M_R$ Model Parameters:} \quad $a' = 1.7,
		\ b = c =   3.4,\ \Lambda_R = 2.5 \times 10^{14}$ GeV}\\[0.1in]
	\multicolumn{7}{c}{\hspace*{0.3in}{\bf Predictions:}\quad 
		$\Delta m^2_{21} = 4.0\times 10^{-5}\ {\rm eV^2},\quad 
		\Delta m^2_{32} = 3.2 \times 10^{-3}\ {\rm eV^2}$}\\[0.1in]
	$\phi'$\hspace*{0.1in} &\hspace*{0.4in} $\sin^2 2\theta_{12}$ 
		&\hspace*{0.4in} $\sin^2 2\theta_{23}$ & \hspace*{0.4in} 
		$\sin^2 2\theta_{13}$ & $\delta_{CP}$\hspace*{0.5in} & 
		$\chi_{1}$ \hspace*{0.5in} & $\chi_{2}$\hspace*{0.10in}\\
		\hline\hline
	$-\pi$ &\hspace*{0.4in} 0.833 &\hspace*{0.5in} 0.18
		&\hspace*{0.5in} 0.0076 & \hspace*{0.4in} $181^\circ$ 
		\hspace*{0.4in} & $-180^\circ$ \hspace*{0.4in} 
		& $-180^\circ$ \\
	$-3\pi/4$ &\hspace*{0.4in} 0.811 &\hspace*{0.5in} 0.38
		&\hspace*{0.5in} 0.0071 & \hspace*{0.4in} $100^\circ$ 
		\hspace*{0.4in} & $-113^\circ$ \hspace*{0.4in} 
		& $-111^\circ$ \\
	$-\pi/2$ &\hspace*{0.4in} 0.793 &\hspace*{0.5in} 0.75
		&\hspace*{0.5in} 0.0057 & \hspace*{0.4in} $59^\circ$ 
		\hspace*{0.4in} & $-74^\circ$ \hspace*{0.4in} 
		& $-70^\circ$ \\
	\hline
	$-\pi/3$ &\hspace*{0.4in} 0.774 &\hspace*{0.5in} 0.916
		&\hspace*{0.5in} 0.0046 & \hspace*{0.4in} $37^\circ$ 
		\hspace*{0.4in} & $-50^\circ$ \hspace*{0.4in} 
		& $-47^\circ$\\ 
	$-\pi/4$ &\hspace*{0.4in} 0.764 &\hspace*{0.5in} 0.958
		&\hspace*{0.5in} 0.0041 & \hspace*{0.4in} $27^\circ$ 
		\hspace*{0.4in} & $-38^\circ$ \hspace*{0.4in} 
		& $-36^\circ$ \\
	$0$ &\hspace*{0.4in} 0.747 &\hspace*{0.5in} 0.992
		&\hspace*{0.5in} 0.0033 & \hspace*{0.4in} $-2^\circ$ 
		\hspace*{0.4in} & $0^\circ$ \hspace*{0.4in} 
		& $0^\circ$\\
	$\pi/4$ &\hspace*{0.4in} 0.757 &\hspace*{0.5in} 0.958
		&\hspace*{0.5in} 0.0038 & \hspace*{0.4in} $-28^\circ$ 
		\hspace*{0.4in} & $38^\circ$ \hspace*{0.4in} 
		& $36^\circ$\\
	$\pi/3$ &\hspace*{0.4in} 0.765 &\hspace*{0.5in} 0.916
		&\hspace*{0.5in} 0.0042 & \hspace*{0.4in} $-38^\circ$ 
		\hspace*{0.4in} & $50^\circ$ \hspace*{0.4in} 
		& $47^\circ$\\
	\hline
	$\pi/2$ &\hspace*{0.4in} 0.784 &\hspace*{0.5in} 0.75
		&\hspace*{0.5in} 0.0052 & \hspace*{0.4in} $-59^\circ$ 
		\hspace*{0.4in} & $74^\circ$ \hspace*{0.4in} 
		& $70^\circ$ \\
	$3\pi/4$ &\hspace*{0.4in} 0.805 &\hspace*{0.5in} 0.38
		&\hspace*{0.5in} 0.0068 & \hspace*{0.4in} $-99^\circ$ 
		\hspace*{0.4in} & $113^\circ$ \hspace*{0.4in} 
		& $111^\circ$\\
	$\pi$ &\hspace*{0.4in} 0.833 &\hspace*{0.5in} 0.18
		&\hspace*{0.5in} 0.0076 & \hspace*{0.4in} $-179^\circ$ 
		\hspace*{0.4in} & $180^\circ$ \hspace*{0.4in} 
		& $180^\circ$\\
\end{tabular}
\end{table}

\begin{table}\label{tab:x12}
\caption[]{Oscillation Parameters for Point (F)\\}
\begin{tabular}{rcllrrr}
	\multicolumn{7}{l}{{\bf (F) $M_R$ Model Parameters:} \quad $a' = 1.0,
		\ b = c = 
		3.0,\ \Lambda_R = 2.6 \times 10^{14}$ GeV}\\[0.1in]
	\multicolumn{7}{c}{\hspace*{0.3in}{\bf Predictions:}\quad 
		$\Delta m^2_{21} = 12.8\times 10^{-5}\ {\rm eV^2},\quad 
		\Delta m^2_{32} = 3.2 \times 10^{-3}\ {\rm eV^2}$}\\[0.1in]
	$\phi'$\hspace*{0.1in} &\hspace*{0.4in} $\sin^2 2\theta_{12}$ 
		&\hspace*{0.4in} $\sin^2 2\theta_{23}$ & \hspace*{0.4in} 
		$\sin^2 2\theta_{13}$ & $\delta_{CP} $\hspace*{0.5in} & 
		$\chi_{1}$ \hspace*{0.5in} & $\chi_{2}$\hspace*{0.10in}\\
		\hline\hline
	$-\pi$ &\hspace*{0.4in} 0.750 &\hspace*{0.5in} 0.06
		&\hspace*{0.5in} 0.0018 & \hspace*{0.4in} $183^\circ$ 
		\hspace*{0.4in} & $-180^\circ$ \hspace*{0.4in} 
		& $-180^\circ$\\
	$-3\pi/4$ &\hspace*{0.4in} 0.736 &\hspace*{0.5in} 0.34
		&\hspace*{0.5in} 0.0016 & \hspace*{0.4in} $76^\circ$ 
		\hspace*{0.4in} & $-99^\circ$ \hspace*{0.4in} 
		& $-96^\circ$\\
	$-\pi/2$ &\hspace*{0.4in} 0.710 &\hspace*{0.5in} 0.82
		&\hspace*{0.5in} 0.0010 & \hspace*{0.4in} $27^\circ$ 
		\hspace*{0.4in} & $-68^\circ$ \hspace*{0.4in} 
		& $-65^\circ$ \\
	\hline
	$-\pi/3$ &\hspace*{0.4in} 0.685 &\hspace*{0.5in} 0.975
		&\hspace*{0.5in} 0.0005 & \hspace*{0.4in} $-5^\circ$ 
		\hspace*{0.4in} & $-48^\circ$ \hspace*{0.4in} 
		& $-45^\circ$\\ 
	$-\pi/4$ &\hspace*{0.4in} 0.674 &\hspace*{0.5in} 0.998
		&\hspace*{0.5in} 0.0003 & \hspace*{0.4in} $-22^\circ$ 
		\hspace*{0.4in} & $-38^\circ$ \hspace*{0.4in} 
		& $-35^\circ$ \\
	$0$ &\hspace*{0.4in} 0.655 &\hspace*{0.5in} 0.987
		&\hspace*{0.5in} 0.00001 & \hspace*{0.4in} $-50^\circ$ 
		\hspace*{0.4in} & $0^\circ$  \hspace*{0.4in} 
		& $0^\circ$\\
	$\pi/4$ &\hspace*{0.4in} 0.665 &\hspace*{0.5in} 0.998
		&\hspace*{0.5in} 0.0002 & \hspace*{0.4in} $27^\circ$ 
		\hspace*{0.4in} & $38^\circ$ \hspace*{0.4in} 
		& $35^\circ$ \\
	$\pi/3$ &\hspace*{0.4in} 0.675 &\hspace*{0.5in} 0.975
		&\hspace*{0.5in} 0.0004 & \hspace*{0.4in} $11^\circ$ 
		\hspace*{0.4in} & $48^\circ$ \hspace*{0.4in} 
		& $45^\circ$\\
	\hline
	$\pi/2$ &\hspace*{0.4in} 0.698 &\hspace*{0.5in} 0.82
		&\hspace*{0.5in} 0.0008 & \hspace*{0.4in} $-21^\circ$ 
		\hspace*{0.4in} & $68^\circ$ \hspace*{0.4in} 
		& $65^\circ$ \\
	$3\pi/4$ &\hspace*{0.4in} 0.729 &\hspace*{0.5in} 0.34
		&\hspace*{0.5in} 0.0014 & \hspace*{0.4in} $-71^\circ$ 
		\hspace*{0.4in} & $99^\circ$ \hspace*{0.4in} 
		& $96^\circ$\\
	$\pi$ &\hspace*{0.4in} 0.750 &\hspace*{0.5in} 0.06
		&\hspace*{0.5in} 0.0018 & \hspace*{0.4in} $-177^\circ$ 
		\hspace*{0.4in} & $180^\circ$ \hspace*{0.4in} 
		& $180^\circ$\\
\end{tabular}
\end{table}

\begin{table}\label{tab:x13}
\caption[]{Oscillation Parameters for Point (G)\\}
\begin{tabular}{rcllrrr}
	\multicolumn{7}{l}{{\bf (G) $M_R$ Model Parameters:} \quad $a' = 1.3,
		\ b = c = 
		3.5,\ \Lambda_R = 2.4 \times 10^{14}$ GeV}\\[0.1in]
	\multicolumn{7}{c}{\hspace*{0.3in}{\bf Predictions:}\quad 
		$\Delta m^2_{21} = 8.8\times 10^{-5}\ {\rm eV^2},\quad 
		\Delta m^2_{32} = 3.2 \times 10^{-3}\ {\rm eV^2}$}\\[0.1in]
	$\phi'$\hspace*{0.1in} &\hspace*{0.4in} $\sin^2 2\theta_{12}$ 
		&\hspace*{0.4in} $\sin^2 2\theta_{23}$ & \hspace*{0.4in} 
		$\sin^2 2\theta_{13}$ & $\delta_{CP}$\hspace*{0.5in} & 
		$\chi_{1}$ \hspace*{0.5in} & $\chi_{2}$\hspace*{0.10in}\\
		\hline\hline
	$-\pi$ &\hspace*{0.4in} 0.726 &\hspace*{0.5in} 0.08
		&\hspace*{0.5in} 0.0043 & \hspace*{0.4in} $182^\circ$ 
		\hspace*{0.4in} & $-180^\circ$ \hspace*{0.4in} 
		& $-180^\circ$\\
	$-3\pi/4$ &\hspace*{0.4in} 0.706 &\hspace*{0.5in} 0.35
		&\hspace*{0.5in} 0.0039 & \hspace*{0.4in} $87^\circ$ 
		\hspace*{0.4in} & $-104^\circ$ \hspace*{0.4in} 
		& $-101^\circ$ \\
	$-\pi/2$ &\hspace*{0.4in} 0.684 &\hspace*{0.5in} 0.80
		&\hspace*{0.5in} 0.0027 & \hspace*{0.4in} $47^\circ$ 
		\hspace*{0.4in} & $-70^\circ$ \hspace*{0.4in} 
		& $-66^\circ$\\
	\hline
	$-\pi/3$ &\hspace*{0.4in} 0.659 &\hspace*{0.5in} 0.961
		&\hspace*{0.5in} 0.0019 & \hspace*{0.4in} $25^\circ$ 
		\hspace*{0.4in} & $-49^\circ$ \hspace*{0.4in} 
		& $-46^\circ$\\ 
	$-\pi/4$ &\hspace*{0.4in} 0.648 &\hspace*{0.5in} 0.991
		&\hspace*{0.5in} 0.0015 & \hspace*{0.4in} $15^\circ$ 
		\hspace*{0.4in} & $-38^\circ$ \hspace*{0.4in} 
		& $-35^\circ$ \\
	$0$ &\hspace*{0.4in} 0.629 &\hspace*{0.5in} 0.996
		&\hspace*{0.5in} 0.0008 & \hspace*{0.4in} $-4^\circ$ 
		\hspace*{0.4in} & $0^\circ$ \hspace*{0.4in} 
		& $0^\circ$\\
	$\pi/4$ &\hspace*{0.4in} 0.639 &\hspace*{0.5in} 0.991
		&\hspace*{0.5in} 0.0012 & \hspace*{0.4in} $-17^\circ$ 
		\hspace*{0.4in} & $38^\circ$ \hspace*{0.4in} 
		& $35^\circ$\\
	$\pi/3$ &\hspace*{0.4in} 0.649 &\hspace*{0.5in} 0.961
		&\hspace*{0.5in} 0.0015 & \hspace*{0.4in} $-25^\circ$ 
		\hspace*{0.4in} & $49^\circ$ \hspace*{0.4in} 
		& $46^\circ$\\
	\hline
	$\pi/2$ &\hspace*{0.4in} 0.672 &\hspace*{0.5in} 0.80
		&\hspace*{0.5in} 0.0023 & \hspace*{0.4in} $-46^\circ$ 
		\hspace*{0.4in} & $70^\circ$ \hspace*{0.4in} 
		& $66^\circ$ \\
	$3\pi/4$ &\hspace*{0.4in} 0.699 &\hspace*{0.5in} 0.35
		&\hspace*{0.5in} 0.0036 & \hspace*{0.4in} $-85^\circ$ 
		\hspace*{0.4in} & $104^\circ$ \hspace*{0.4in} 
		& $101^\circ$\\
	$\pi$ &\hspace*{0.4in} 0.726 &\hspace*{0.5in} 0.08
		&\hspace*{0.5in} 0.0043 & \hspace*{0.4in} $-178^\circ$ 
		\hspace*{0.4in} & $180^\circ$ \hspace*{0.4in} 
		& $180^\circ$\\
\end{tabular}
\end{table}

\begin{table}\label{tab:x14}
\caption[]{Oscillation Parameters for Point (H)\\}
\begin{tabular}{rcllrrr}
	\multicolumn{7}{l}{{\bf (H) $M_R$ Model Parameters:} \quad $a' = 2.1,
		\ b = c = 
		4.3,\ \Lambda_R = 2.4 \times 10^{14}$ GeV}\\[0.1in]
	\multicolumn{7}{c}{\hspace*{0.3in}{\bf Predictions:}\quad 
		$\Delta m^2_{21} = 3.6\times 10^{-5}\ {\rm eV^2},\quad 
		\Delta m^2_{32} = 3.2 \times 10^{-3}\ {\rm eV^2}$}\\[0.1in]
	$\phi'$\hspace*{0.1in} &\hspace*{0.4in} $\sin^2 2\theta_{12}$ 
		&\hspace*{0.4in} $\sin^2 2\theta_{23}$ & \hspace*{0.4in} 
		$\sin^2 2\theta_{13}$ & $\delta_{CP}$\hspace*{0.5in} & 
		$\chi_{1}$ \hspace*{0.5in} & $\chi_{2}$\hspace*{0.10in}\\
		\hline\hline
	$-\pi$ &\hspace*{0.4in} 0.746 &\hspace*{0.5in} 0.17
		&\hspace*{0.5in} 0.0090 & \hspace*{0.4in} $181^\circ$ 
		\hspace*{0.4in} & $-180^\circ$ \hspace*{0.4in} 
		& $-180^\circ$\\
	$-3\pi/4$ &\hspace*{0.4in} 0.719 &\hspace*{0.5in} 0.38
		&\hspace*{0.5in} 0.0084 & \hspace*{0.4in} $101^\circ$ 
		\hspace*{0.4in} & $-113^\circ$ \hspace*{0.4in} 
		& $-111^\circ$\\
	$-\pi/2$ &\hspace*{0.4in} 0.699 &\hspace*{0.5in} 0.76
		&\hspace*{0.5in} 0.0068 & \hspace*{0.4in} $60^\circ$ 
		\hspace*{0.4in} & $-74^\circ$ \hspace*{0.4in} 
		& $-70^\circ$ \\
	\hline
	$-\pi/3$ &\hspace*{0.4in} 0.678 &\hspace*{0.5in} 0.919
		&\hspace*{0.5in} 0.0056 & \hspace*{0.4in} $38^\circ$ 
		\hspace*{0.4in} & $-50^\circ$ \hspace*{0.4in} 
		& $-47^\circ$\\ 
	$-\pi/4$ &\hspace*{0.4in} 0.667 &\hspace*{0.5in} 0.960
		&\hspace*{0.5in} 0.0051 & \hspace*{0.4in} $28^\circ$ 
		\hspace*{0.4in} & $-39^\circ$ \hspace*{0.4in} 
		& $-36^\circ$ \\
	$0$ &\hspace*{0.4in} 0.648 &\hspace*{0.5in} 0.993
		&\hspace*{0.5in} 0.0042 & \hspace*{0.4in} $-1^\circ$ 
		\hspace*{0.4in} & $0^\circ$  \hspace*{0.4in} 
		& $0^\circ$\\
	$\pi/4$ &\hspace*{0.4in} 0.659 &\hspace*{0.5in} 0.960
		&\hspace*{0.5in} 0.0047 & \hspace*{0.4in} $-29^\circ$ 
		\hspace*{0.4in} & $38^\circ$ \hspace*{0.4in} 
		& $36^\circ$ \\
	$\pi/3$ &\hspace*{0.4in} 0.667 &\hspace*{0.5in} 0.919
		&\hspace*{0.5in} 0.0051 & \hspace*{0.4in} $-39^\circ$ 
		\hspace*{0.4in} & $50^\circ$ \hspace*{0.4in} 
		& $47^\circ$\\
	\hline
	$\pi/2$ &\hspace*{0.4in} 0.688 &\hspace*{0.5in} 0.76
		&\hspace*{0.5in} 0.0063 & \hspace*{0.4in} $-60^\circ$ 
		\hspace*{0.4in} & $74^\circ$ \hspace*{0.4in} 
		& $70^\circ$ \\
	$3\pi/4$ &\hspace*{0.4in} 0.712 &\hspace*{0.5in} 0.38
		&\hspace*{0.5in} 0.0080 & \hspace*{0.4in} $-99^\circ$ 
		\hspace*{0.4in} & $113^\circ$ \hspace*{0.4in} 
		& $111^\circ$\\
	$\pi$ &\hspace*{0.4in} 0.746 &\hspace*{0.5in} 0.17
		&\hspace*{0.5in} 0.0090 & \hspace*{0.4in} $-179^\circ$ 
		\hspace*{0.4in} & $180^\circ$ \hspace*{0.4in} 
		& $180^\circ$\\
\end{tabular}
\end{table}

\end{document}